\documentclass[ALICE,manyauthors]{cernphprep}
\usepackage[comma,square,numbers,sort&compress]{natbib}

\usepackage{lineno}
\usepackage{xspace}
\usepackage{hyperref}
\usepackage{xcolor}
\usepackage{overpic} 

\usepackage[T1]{fontenc}
\usepackage{orcidlink} 

\newcommand{\he}{\ensuremath{{}^3\mathrm{He}}}

\newcommand{\hyp}{\ensuremath{{}^3_{\Lambda}\mathrm{H}}}
\newcommand{\ahyp}{\ensuremath{{}^3_{\bar{\Lambda}}\overline{\mathrm{H}}}}
\newcommand{\hypToL}{\ensuremath{{}^3_{\Lambda}\mathrm{H}/\Lambda}}

\newcommand{\rd}{\ensuremath{\sqrt{\left<r^2_\mathrm{\mathrm{d}}\right>}}\xspace}

\newcommand{\rhe}{\ensuremath{\sqrt{\left< r^2_{{}^{3}\mathrm{He}}\right>}}\xspace}

\newcommand{\rhyp}{\ensuremath{\sqrt{\left<r^2_{\mathrm{^3_{\Lambda}H}}\right>}}\xspace}
\newcommand{\rrhyp}{\ensuremath{\sqrt{\left<r^2_{\mathrm{d \Lambda}}\right>}}\xspace}

\newcommand{\mt}           {\ensuremath{m_{\rm T}}\xspace}

\begin{document}
%%%%%%%%%%%%%%%%%%%%%%%%%%%%%%%%%%%%%%%%%%%%%%%%%%
% These are some new commands that may be useful 
% for paper writing in general. If other newcommands
% are needed for your specific paper, please feel 
% free to add here. 
%
% The currently available commands are organized in: 
% 1) Systems
% 2) Quantities
% 3) Energies and units
% 4) Detectors
% 5) particle species 
%%%%%%%%%%%%%%%%%%%%%%%%%%%%%%%%%%%%%%%%%%%%%%%%%%

% 1) SYSTEMS 
\newcommand{\pp}           {pp\xspace}
\newcommand{\ppbar}        {\mbox{$\mathrm {p\overline{p}}$}\xspace}
\newcommand{\XeXe}         {\mbox{Xe--Xe}\xspace}
\newcommand{\PbPb}         {\mbox{Pb--Pb}\xspace}
\newcommand{\pA}           {\mbox{pA}\xspace}
\newcommand{\pPb}          {\mbox{p--Pb}\xspace}
\newcommand{\AuAu}         {\mbox{Au--Au}\xspace}
\newcommand{\dAu}          {\mbox{d--Au}\xspace}

% 2) QUANTITIES 
\newcommand{\s}            {\ensuremath{\sqrt{s}}\xspace}
\newcommand{\snn}          {\ensuremath{\sqrt{s_{\mathrm{NN}}}}\xspace}
\newcommand{\pt}           {\ensuremath{p_{\rm T}}\xspace}
\newcommand{\meanpt}       {$\langle p_{\mathrm{T}}\rangle$\xspace}
\newcommand{\ycms}         {\ensuremath{y_{\rm CMS}}\xspace}
\newcommand{\ylab}         {\ensuremath{y_{\rm lab}}\xspace}
\newcommand{\etarange}[1]  {\mbox{$\left | \eta \right |~<~#1$}}
\newcommand{\yrange}[1]    {\mbox{$\left | y \right |~<~#1$}}
\newcommand{\dndy}         {\ensuremath{\mathrm{d}N_\mathrm{ch}/\mathrm{d}y}\xspace}
\newcommand{\dndeta}       {\ensuremath{\mathrm{d}N_\mathrm{ch}/\mathrm{d}\eta}\xspace}
\newcommand{\avdndeta}     {\ensuremath{\langle\dndeta\rangle_{|\eta|<0.5}}\xspace}
\newcommand{\dNdy}         {\ensuremath{\mathrm{d}N_\mathrm{ch}/\mathrm{d}y}\xspace}
\newcommand{\Npart}        {\ensuremath{N_\mathrm{part}}\xspace}
\newcommand{\Ncoll}        {\ensuremath{N_\mathrm{coll}}\xspace}
\newcommand{\dEdx}         {\ensuremath{\textrm{d}E/\textrm{d}x}\xspace}
\newcommand{\RpPb}         {\ensuremath{R_{\rm pPb}}\xspace}

% 3) ENERGIES, UNITS
\newcommand{\nineH}        {$\sqrt{s}~=~0.9$~Te\kern-.1emV\xspace}
\newcommand{\seven}        {$\sqrt{s}~=~7$~Te\kern-.1emV\xspace}
\newcommand{\twoH}         {$\sqrt{s}~=~0.2$~Te\kern-.1emV\xspace}
\newcommand{\twosevensix}  {$\sqrt{s}~=~2.76$~Te\kern-.1emV\xspace}
\newcommand{\five}         {$\sqrt{s}~=~5.02$~Te\kern-.1emV\xspace}
\newcommand{\twosevensixnn}{$\sqrt{s_{\mathrm{NN}}}~=~2.76$~Te\kern-.1emV\xspace}
\newcommand{\fivenn}       {$\sqrt{s_{\mathrm{NN}}}~=~5.02$~Te\kern-.1emV\xspace}
\newcommand{\LT}           {L{\'e}vy-Tsallis\xspace}
\newcommand{\GeVc}         {Ge\kern-.1emV/$c$\xspace}
\newcommand{\MeVc}         {Me\kern-.1emV/$c$\xspace}
\newcommand{\TeV}          {Te\kern-.1emV\xspace}
\newcommand{\GeV}          {Ge\kern-.1emV\xspace}
\newcommand{\MeV}          {Me\kern-.1emV\xspace}
\newcommand{\GeVmass}      {Ge\kern-.2emV/$c^2$\xspace}
\newcommand{\MeVmass}      {Me\kern-.2emV/$c^2$\xspace}
\newcommand{\lumi}         {\ensuremath{\mathcal{L}}\xspace}

% 4) DETECTORS 
\newcommand{\ITS}          {\rm{ITS}\xspace}
\newcommand{\TOF}          {\rm{TOF}\xspace}
\newcommand{\ZDC}          {\rm{ZDC}\xspace}
\newcommand{\ZDCs}         {\rm{ZDCs}\xspace}
\newcommand{\ZNA}          {\rm{ZNA}\xspace}
\newcommand{\ZNC}          {\rm{ZNC}\xspace}
\newcommand{\SPD}          {\rm{SPD}\xspace}
\newcommand{\SDD}          {\rm{SDD}\xspace}
\newcommand{\SSD}          {\rm{SSD}\xspace}
\newcommand{\TPC}          {\rm{TPC}\xspace}
\newcommand{\TRD}          {\rm{TRD}\xspace}
\newcommand{\VZERO}        {\rm{V0}\xspace}
\newcommand{\VZEROA}       {\rm{V0A}\xspace}
\newcommand{\VZEROC}       {\rm{V0C}\xspace}
\newcommand{\Vdecay} 	   {\ensuremath{V^{0}}\xspace}

% 4) PARTICLE SPECIES 
\newcommand{\ee}           {\ensuremath{e^{+}e^{-}}} 
\newcommand{\pip}          {\ensuremath{\pi^{+}}\xspace}
\newcommand{\pim}          {\ensuremath{\pi^{-}}\xspace}
\newcommand{\kap}          {\ensuremath{\rm{K}^{+}}\xspace}
\newcommand{\kam}          {\ensuremath{\rm{K}^{-}}\xspace}
\newcommand{\pbar}         {\ensuremath{\rm\overline{p}}\xspace}
\newcommand{\kzero}        {\ensuremath{{\rm K}^{0}_{\rm{S}}}\xspace}
\newcommand{\lmb}          {\ensuremath{\Lambda}\xspace}
\newcommand{\almb}         {\ensuremath{\overline{\Lambda}}\xspace}
\newcommand{\Om}           {\ensuremath{\Omega^-}\xspace}
\newcommand{\Mo}           {\ensuremath{\overline{\Omega}^+}\xspace}
\newcommand{\X}            {\ensuremath{\Xi^-}\xspace}
\newcommand{\Ix}           {\ensuremath{\overline{\Xi}^+}\xspace}
\newcommand{\Xis}          {\ensuremath{\Xi^{\pm}}\xspace}
\newcommand{\Oms}          {\ensuremath{\Omega^{\pm}}\xspace}
\newcommand{\degree}       {\ensuremath{^{\rm o}}\xspace}

%%%%%%%%%%%%%%%  Title page %%%%%%%%%%%%%%%%%%%%%%%%
\begin{titlepage}
% the dates below correspond to CERN approval
% please don't touch: EB chairs will take care
\PHyear{2026}       % required, will be obtained from CERN
\PHnumber{097}      % required, will be obtained from CERN
\PHdate{23 March}  % required, will be obtained from CERN
%%%%%%%%%%%%%%%%%%%%%%%%%%%%%%%%%%%%%%%%%%%%%%%%%%%%

%%% Put your own title + short title here:
\title{Wave-Function Femtometry:\\Hypertriton - The Ultimate Halo Nucleus}
\ShortTitle{Wave-function femtometry}   % appears on left page headers

%%% Do not change the next lines
\Collaboration{ALICE Collaboration\thanks{See Appendix~\ref{app:collab} for the list of collaboration members}}
\ShortAuthor{ALICE Collaboration} % appears on right page headers, do not change

\begin{abstract}

The interaction between nucleons and hyperons -- baryons containing a strange quark -- is key to understanding the properties of dense nuclear matter, such as that expected in the interior of neutron stars. Direct scattering experiments are hindered by the short lifetime of hyperons, prompting the study of hypernuclei -- bound states of nucleons and hyperons -- as an alternative approach.  The lightest known hypernucleus, the hypertriton (\hyp), is a weakly bound state composed of a proton, a neutron and a $\Lambda$ hyperon, and is believed to exhibit a halo-like structure with the $\Lambda$ being loosely bound to a deuteron core. Based on the first measurement of hypertriton production in proton-proton collisions at the CERN Large Hadron Collider (LHC), its halo structure is confirmed. A successful description of the hypertriton production yield within the nuclear coalescence framework enables an
estimation of the $\Lambda$ separation from the deuteron core as  $9.54^{+2.67}_{-1.11}$ fm. 

\end{abstract}
\end{titlepage}

\setcounter{page}{2} %please do not remove this line

%%%%%%%%%%%%%%%%%%%%%%%%%%%%%%%%
% begin main text
%%%%%%%%%%%%%%%%%%%%%%%%%%%%%%%%

%Introduction 
\section{Introduction} 
Protons and neutrons -- the building blocks of atomic nuclei -- are composed of three light valence quarks of up ({\em u}) and down ({\em d}) flavor. The forces that act between protons and neutrons have been explored for more than 100 years, by performing scattering experiments and studying the properties of nuclei, and a remarkable level of understanding has been achieved~\cite{Krane:1987ky,Obertelli:2021crc}. Under extreme conditions of high pressure, such as those in the interior of neutron stars, states of nuclear matter are predicted in which, in addition to protons and neutrons, hyperons can play a role~\cite{Tolos:2020aln,Burgio:2021vgk,Vidana:2024ngv}. Hyperons are baryons that contain at least one strange valence quark ({\em s}). The lightest hyperon is the $\Lambda$ with quark content {\em uds} and a rest mass of [1115.683 $\pm$ 0.006]~MeV/$c^2$~\cite{ParticleDataGroup:2022pth}. The properties of high-density matter containing hyperons are largely determined by the strong interaction of hyperons and nucleons (YN) and hyperons with each other (YY)~\cite{Tolos:2020aln,Burgio:2021vgk}. Because hyperons decay via the weak interaction with lifetimes below a nanosecond, scattering experiments are very challenging, and data are scarce~\cite{Tolos:2020aln,Haidenbauer:2023qhf,Mihaylov:2023ahn}. Recently, new constraints on the YN and YY interactions have been obtained from femtoscopic measurements in production experiments at high collision energy (see for instance~\cite{ALICE:2019eol,ALICE:2020mfd,Mihaylov:2023ahn}). A complementary approach involves the study of hypernuclei, short-lived bound states of hyperons and nucleons. These systems offer unique insights into hypernuclear structure and the underlying baryonic interactions~\cite{RevMod88:Strangeness}.

The lightest known hypernucleus is the hypertriton (\hyp), a bound state of a proton, a neutron, and a $\Lambda$ hyperon. The hypertriton is often treated as a weakly bound state of a deuteron and a $\Lambda$ hyperon. The $\Lambda$ binding energy relative to the proton–neutron subsystem -- conceptualized 
as a deuteron -- has a world average value of $105^{+37}_{-28}$~keV~\cite{HypernuclearDataBase}, more than an order of magnitude smaller than typical nucleon separation energies in ordinary nuclei.
Thus, the quoted value corresponds to the $\Lambda$ separation energy. The total binding energy is approximately 2.32~MeV, given by $B_{\hyp} = B_\Lambda + B_{\mathrm{d}}$, with $B_{\mathrm{d}} = 2.22$~MeV. This suggests that the hypertriton is a so-called halo nucleus ~\cite{Hammer:2001ng,Jensen:2004zz,Braun-Munzinger:2018hat}, an exotic system in which a few nucleons occupy orbitals extending far beyond the compact core~\cite{Hansen:1987mc,Riisager:1994zz,Tanihata:1995yv,Jensen:2004zz,Tanihata:2020atp}. 
A prominent halo nucleus is ${}^{11}$Li, which has a ${}^{9}$Li core with a two-neutron halo. The rms radius of the core is about 3~fm and the one of the halo is about 4~fm~\cite{Tanihata:1995yv}. In the case of the hypertriton, the wave function allows the weakly bound $\Lambda$ to have a high probability of being outside the classically allowed region, which is defined by the short range of the strong interaction of about 1 fm. This assumption is supported by the hypertriton's lifetime of 
$[253 \pm 11~({\rm stat.}) \pm 6~({\rm syst.})]$ ps~\cite{ALICE:2022sco} which, after long-standing experimental controversies~\cite{Prem:1964hyp,Keyes:1968zz,Phillips:1969uy,Bohm:1970se,Keyes:1970ck,Keyes:1974ev,STAR:2010gyg,Rappold:2013fic,ALICE:2015oer,Adamczyk:2017buv,ALICE:2019vlx,ALICE:2022sco}, is now consistent with that of the free $\Lambda$ [263.2 $\pm$ 2.0] ps~\cite{ParticleDataGroup:2022pth}.

Despite strong theoretical indications~\cite{Cobis:1996ru,Nemura:1999qp,Braun-Munzinger:2018hat,Hildenbrand:2019sgp,Bertulani:2022vad}, direct experimental evidence for the halo structure of the hypertriton, such as a measurement of its matter radius, is still lacking. Established techniques, including scattering experiments and laser spectroscopy~\cite{Skawran:2019krq,Grinin:2020txk,Krauth:2021foz} are currently infeasible due to the hypertriton's short lifetime.  However, measurements of its geometric cross section via scattering on a secondary target nucleus are in preparation and expected to provide the first direct estimates of its matter radius~\cite{Velardita:2023lbi}. Consequently,
present knowledge relies on theoretical models, particularly effective field theory (EFT) calculations, which predict a strong anti-correlation between binding energy and spatial extent. Assuming a $\Lambda$ separation energy of 130 keV~\cite{Juric:1973zq}, state-of-the-art calculations yield a root-mean-square (rms) distance  between the deuteron core and $\Lambda$ of about 10 fm~\cite{Nemura:1999qp,Braun-Munzinger:2018hat,Hildenbrand:2019sgp,Hildenbrand:2023rkc}, exceeding even the radius of a heavy nucleus such as lead ($\approx5.5$ fm~\cite{Angeli:2013epw}).

This article follows a complementary approach to determine the matter radius of the hypertriton. It exploits the fact that, within the coalescence model (CM), the production rate of light nuclei and hypernuclei in high-energy nucleus–nucleus or hadron–hadron collisions depends on the wave function of the nuclear cluster to be formed -- in this case, the hypertriton. Generally, the coalescence model assumes that the nucleons, i.e. protons and neutrons, can form a certain nucleus if these nucleons are close in phase space~\cite{butler_pearson61,Butler:1963pp}. The probability of forming such a cluster increases when the phase-space configuration of the produced nucleons and hyperons overlaps with the configuration described by the nuclear wave function. As a result, the sensitivity on the wave function is expected to be particularly pronounced in small collision systems, such as proton–proton (pp), where phase-space densities are lower than in nucleus-nucleus collisions and more sensitive to structural details of the cluster. The phase-space configuration itself is directly related to the system size, which in turn scales with the event’s charged-particle multiplicity.

However, the range of validity of the coalescence picture for describing the production of (hyper-)nuclear clusters is not yet fully established. A recent analysis revealed the validity of the picture in pp collisions~\cite{ALICE:2025zzg}. There are different implementations of the coalescence model, that differ in particular in the particle-emitting source description and the wave function of the nucleus. For instance, the source can be modeled using correlation measurements of pions~\cite{Scheibl:1998tk,Bellini:2018epz} or of protons~\cite{Blum:2017iwq,Bellini:2020cbj,Mahlein:2023fmx,Mahlein:2024pur}, where the source sizes of the systems are about 1--5 fm. The wave function of the nucleus can be approximated by Gaussian wave packets~\cite{Scheibl:1998tk, Sun:2018mqq} or treated with more realistic approaches, including effective chiral modeling~\cite{Bellini:2020cbj,Mahlein:2023fmx,Mahlein:2025bla,Leung:2025jwe}. (The referenced studies show that, without accounting for event-by-event multiplicity fluctuations and momentum correlations, the spatial structure of nuclei -- particularly at low multiplicities -- cannot be reliably constrained, as demonstrated for the deuteron~\cite{ALICE:2021mfm}. A full systematic evaluation of wave-function effects requires 
both more differential data (e.g.\,finely binned momentum spectra) and further development of state-of-the-art theoretical models.) The use of a simplified Gaussian wave function permits an analytical treatment, replacing the need for complex Monte Carlo simulations of the coalescence process. 
In central lead--lead (Pb--Pb) collisions, i.e., large collision systems, statistical hadronization models (SHM), which rely on a grand-canonical description of a thermalized medium, provide an excellent description of the production rates of hadrons, and a reasonably good description of light nuclear clusters~\cite{Andronic:2010qu,Andronic:2017pug,Braun-Munzinger:2018hat,Donigus:2020ctf,Donigus:2022haq}. The most notable discrepancy has recently been observed for hypertriton production in Pb--Pb collisions, where the measured
yield lies significantly below the SHM expectations~\cite{ALICE:2024koa}. In small collision systems such as pp, the grand canonical approach breaks down, and the SHM is extended via a canonical statistical model (CSM) where charge-like quantum numbers (baryon number, strangeness, electric charge) are conserved exactly rather than on average~\cite{Vovchenko:2018fiy,Donigus:2020ctf,Cleymans:2020fsc,Sharma:2022poi}. Predictions from SHM/CSM and coalescence models differ significantly in such systems, making pp collisions an ideal testing ground to study the mechanism of nuclei formation. In particular, 
if the hypertriton exhibits a halo structure, its production is expected to be strongly suppressed in the coalescence model, whereas in the CSM it depends only on the mass and quantum numbers, but not on the spatial size of the nucleus. Recent measurements of hypertriton production in p--Pb collisions are consistent with coalescence calculations and already exclude some part of the CSM parameter space~\cite{ALargeIonColliderExperiment:2021puh}.

In this article, we demonstrate how the validity of the coalescence picture for hypernuclei in small
 collision systems can be substantiated using the first measurement of hypertriton production in pp collisions carried out with the ALICE experiment at the LHC. We show that an analysis of the wave function of nuclear clusters -- referred to here as {\em wave-function femtometry} -- can be performed within the coalescence framework. Using this model-dependent approach,
and assuming a Gaussian form for the wave function, the matter radius of the hypertriton can be extracted. By further exploiting the predicted correlation between matter radius and $\Lambda$ separation energy in theoretical calculations~\cite{Hammer:2001ng,Hildenbrand:2019sgp}, the $\Lambda$ separation energy of the hypertriton can be determined with a precision 
comparable to that of state-of-the-art mass measurements.

\section{Analysis techniques} 

In the present analysis, the first detection of hypertritons in pp collisions has been achieved. The data were collected in the years 2016, 2017, and 2018 in pp collisions at a center-of-mass energy of $\sqrt{s}=13\,$TeV at the Large Hadron Collider (LHC) with the ALICE (A Large Ion Collider Experiment) apparatus. Detailed information about the design and performance of the ALICE setup can be found in~\cite{Aamodt:2008zz,Abelev:2014ffa}. 
Collision events used in this analysis were selected using specialized triggers. 
Two high-multiplicity (HM) triggers were used to select events with a mean charged-particle multiplicity \avdndeta = $30.8\pm0.4$. (The corrected number of charged particles is normalized to one unit in pseudorapidity $\eta$, where $\eta=-\ln\left [ \tan \left ( \frac{\theta}{2}\right ) \right]$ and $\theta$ is the particle's angle with respect to the beam axis.)
A second event sample with  \avdndeta = $6.9\pm0.1$  has been selected using a dedicated online trigger designed to identify nuclei based on the ionization energy loss in the Transition Radiation Detector (TRD). The associated charged-particle multiplicity of the inspected event sample is the same as for data recorded without a dedicated trigger, hereafter referred to as the minimum-bias (MB) data set. Femtoscopic studies have shown that the multiplicity of charged particles is closely related to the size of the particle-emitting system~\cite{ALICE:2022veq,ALICE:2023zbh,ALICE:2023sjd,Bellini:2018epz,Blum:2017qnn}. 
A systematic study of hypertriton production at different charged-particle multiplicity is therefore ideal to distinguish different production mechanisms.
Further details on the used data samples and the TRD nuclei trigger are given in the Methods part~\ref{methods}.

The hypertriton is reconstructed via the charged two-body weak decay \hyp $\to$ \he\ + \pim\ (and charge conjugates) with a branching ratio of 25\%~\cite{Kamada1595}. In the following we will use the particle name for both \hyp ~and \ahyp, assuming an equality of their properties and also an equal production.
A dedicated algorithm is used that detects two-body decay topologies 
during track reconstruction. Particle hypotheses are assigned to the associated daughter tracks based on 
the measurement of the specific energy loss (\dEdx) 
in the Time Projection Chamber (TPC) detector. The rigidity of the daughter tracks is determined by the measurement of their curvature 
in the homogeneous magnetic field of the ALICE solenoid ($B=0.5\,$T), and a kinematic reconstruction
of the invariant mass is performed using the decay daughter mass and charge hypotheses. 
To reduce the background resulting from combinations of particles that do not originate from a hypertriton decay, topological and kinematic selections are applied.
More information about the reconstruction details is provided in the Methods part~\ref{methods}.

The invariant-mass distributions of the reconstructed \hyp ~and \ahyp ~candidates are added and fitted with a model including a signal and a background component. A Monte Carlo template is used for the signal component, smoothed with a Kernel Density Estimator function~\cite{Cranmer:2000du,Verkerke:2003ir}, and an exponential is applied to model the background. 
The associated significance for the combined \hyp ~and \ahyp ~signal is 9.5$\,\sigma$ in the HM sample and  5.6$\,\sigma$ in the MB sample, obtained from asymptotic likelihood formulas as done in~\cite{Cowan:2010js}.

The raw (uncorrected) signal counts are extracted from fits to the invariant-mass spectra and corrected for the geometric acceptance of the ALICE detectors and reconstruction efficiency using MC simulations. An additional correction of 3\% is applied to account for absorption processes~\cite{ALargeIonColliderExperiment:2021puh}. 
Systematic uncertainties are determined by variation of the selection criteria, including track selection, particle identification and topological and kinematic criteria. Furthermore, different fit functions have been implemented for the signal and background description and additional systematic uncertainties are added due to uncertainties in the material budget, absorption, and the branching ratio. For more details see the Methods part~\ref{methods}.

\begin{figure}[tb]
    \begin{center}
    \includegraphics[width = 0.8\textwidth]{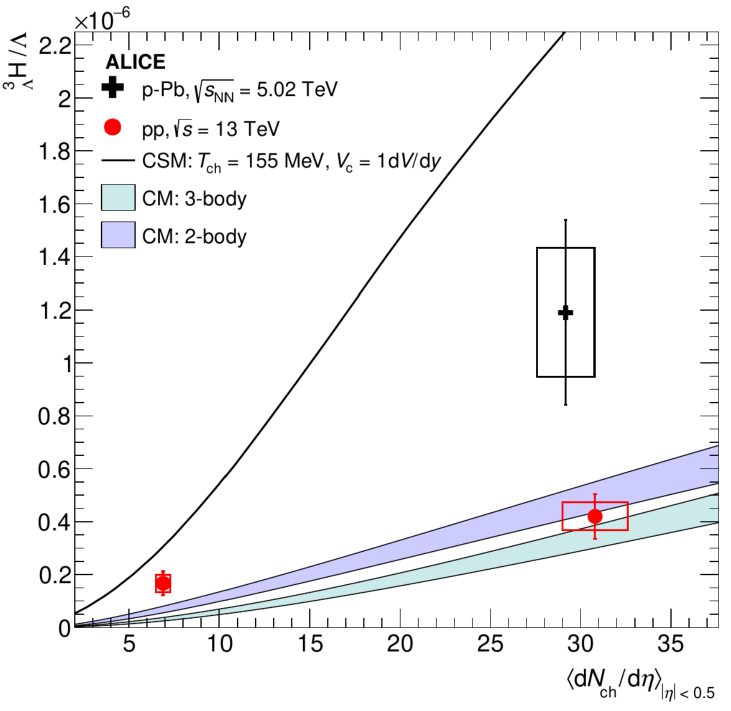}
    \end{center}
    \caption{The \hypToL\ ratio as a function of the mean charged-particle multiplicity (\avdndeta) measured at $\sqrt{s} = 13$ TeV for two multiplicity classes (full red circles), together with a previous experimental result in p--Pb collisions~\cite{ALargeIonColliderExperiment:2021puh} (black cross). 
    Furthermore the CSM thermal model prediction~\cite{Vovchenko:2018fiy,Vovchenko:2019kes} is displayed as black line and the blue and green bands represent the predictions of the two-body and three-body coalescence~\cite{Sun:2018mqq}, respectively. The vertical lines represent the statistical uncertainties, while the vertical boxes are the systematic ones.}
    \label{fig:3LH_L}
\end{figure}

\section{Results}
We observe a hypertriton production yield of d$N$/d$y^{\mathrm{MB}}$ = [2.1 $\pm$ 0.6 (stat.) $\pm$ 0.4 (syst.)]$\times 10^{-8}$ for the MB sample and d$N$/d$y^{\mathrm{HM}}$ = [2.4 $\pm$ 0.5 (stat.) $\pm$ 0.3 (syst.)]$\times 10^{-7}$ for the HM sample. The yield is given as the average of \hyp ~and \ahyp~ and quoted as rapidity density, 
where the rapidity $y$ is related to the relativistic velocity along the beam axis ($y=\frac{1}{2}\ln \frac{E+p_Lc}{E-p_Lc}$, where $E$ is the energy of the particle, $c$ is the speed of light and $p_L$ is the momentum along the beam axis). 

In order to compare the measurements with model predictions, the \hypToL\ ratio has been constructed. The $\Lambda$ yields are taken from previous ALICE measurements of differential $\Lambda$ and $\overline{\Lambda}$ production in pp collisions at the same center-of-mass energy~\cite{ALICE:2019avo,ALICE:2020nkc,ALICE:2020jsh,ALICE:2021mfm}. 
To explore the system size dependence, the \hypToL\ ratios are shown as a function of the charged-particle multiplicity $\langle \dndeta \rangle$ in Fig.~\ref{fig:3LH_L} together with a previous ALICE result in p--Pb~\cite{ALargeIonColliderExperiment:2021puh}  collisions.  
In addition, the predictions of the CSM~\cite{Vovchenko:2018fiy,Vovchenko:2019kes} and the coalescence model implementation from~\cite{Sun:2018mqq} are shown. 

In the case of the CSM, the correlation volume $V_{\rm c}$ is set equal to the volume of the fireball in one unit of rapidity (d$V$/d$y$),
where $V_{\rm c}$ is the volume within which exact conservation of all quantum numbers is enforced. The actual size of $V_{\rm c}$ is 
{\em a priori} unknown~\cite{Castorina:2013mba,Sharma:2022poi} and needs to be constrained by data.  
Nuclei measurements in small collision systems are more consistent with small $V_{\rm c}$~\cite{Vovchenko:2018fiy} while other light-flavor measurements favor \mbox{$V_{\rm c}\approx3$d$V$/d$y$}~\cite{Vovchenko:2019kes,ALICE:2024rnr}.
It should also be noted that different values of $V_{\rm c}$ may apply for 
different conserved quantities.

The curves from the CM include two different assumptions on the inner structure of the hypertriton. The so-called two-body coalescence assumes that the hypertriton is composed of a compact deuteron loosely bound with a $\Lambda$. In this case, the size of the hypertriton is defined as the rms distance between the deuteron and $\Lambda$, \rrhyp. In contrast, the three-body coalescence assumes a composition of a proton, neutron and a $\Lambda$ with equal spatial distribution and the rms matter radius \rhyp is used, which represents the distance from the center-of-mass of the hypertriton to its constituents. For the CM predictions shown in Fig.~\ref{fig:3LH_L}, \rrhyp = 10 fm~\cite{Nemura:1999qp} and \rhyp = 4.9 fm~\cite{Nemura:1999qp} are used for the two-body and three-body coalescence cases, respectively.

The curves of the CSM and the coalescence model are well separated at low charged-particle multiplicities ($\langle \dndeta \rangle \lessapprox 100$) but tend to converge at large $\langle \dndeta \rangle$, where the size of the particle-emitting source approaches that of the hypertriton. In the range of such high multiplicities, other effects become relevant, e.g. absorption of nucleons, which lead to a suppression of the production of clusters of light nuclei, and which are not taken into account in the CSM or the coalescence models. This suppression is predicted for all nuclei studied in UrQMD model calculations~\cite{Reichert:2022mek,ALICE:2022veq,ALICE:2023qyl,ALICE:2024say}.  Therefore, the results measured in Pb–Pb collisions cannot be used to form  a reliable distinction between CSM and the coalescence model (and are therefore not shown). 
However, the measured yields in small collision systems (pp, p--Pb)  favor the (two-body) coalescence picture over the CSM approach. The MB and HM pp data points are $2.4\,\sigma$ and $19.5\,\sigma$ away from the closest CSM curve, respectively, while both points are compatible with the two-body coalescence curve (1.7\,$\sigma$ and 0.6\,$\sigma$). Below we argue that this finding opens the door for a novel technique, which we call {\em wave-function femtometry}, to study the wave function of composite objects by measuring their production rate in small collision systems and applying the coalescence approach to determine the size of the object.

\section{Determination of the hypertriton radius and the $\Lambda$ separation energy}

In the analytical coalescence model, the suppression of the production yield of a composite object depends on the ratio of the object size to the source size. The two-body coalescence approach accounts for the weak binding of the $\Lambda$  and is therefore used to estimate the size of the hypertriton. The \hypToL\ yield ratio in this  approach~\cite{Sun:2018mqq} is given by
\begin{equation}
\label{coal2_Hyp}
\frac{{}^3_{\Lambda}\mathrm{H}}{\Lambda} = \frac{7.1 \times 10^{-6} \times 0.85}{\left[ 1 + \biggl( \sqrt{\frac{2}{9}}\cdot \frac{\rrhyp}{R} \biggr)^2 \right]^{3/2} \left[ 1 + \biggl( \frac{3.2}{2R} \biggr)^2 \right]^{3/2},}
\end{equation}
where $R$ is the source size which can be related to the charged-particle multiplicity. 
This relation opens a new possibility to determine the size of a composite object from the measured particle ratios. 

\begin{figure}[tb]
    \begin{center}
    \includegraphics[width = 1.0\textwidth]{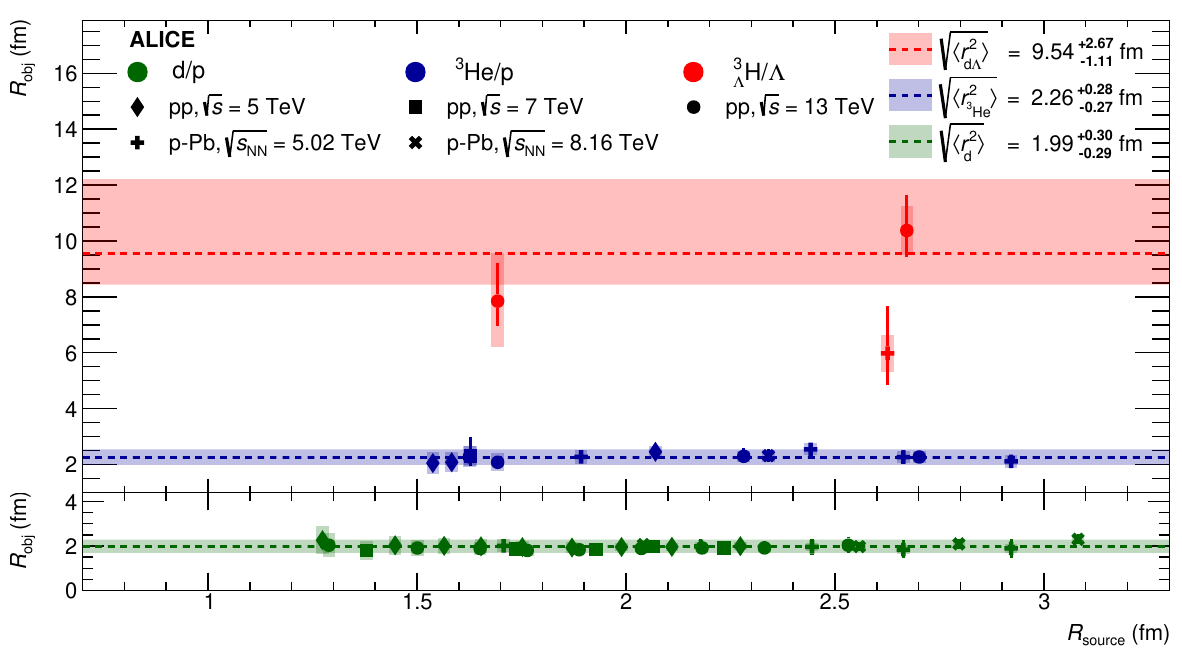}
    \end{center}
    \caption{Object sizes of deuteron (green), \he\ (blue) and \hyp\ (red) obtained from experimental d/p~\cite{ALICE:2019bnp,ALICE:2019dgz,ALICE:2020foi,ALICE:2021ovi}, \he/p~\cite{ALICE:2017xrp,ALICE:2021ovi,ALICE:2021mfm} and \hyp/$\Lambda$ ratios using the corresponding coalescence formulae.  The vertical bars represent the statistical uncertainty resulting from the measured yield ratio. The shaded boxes show the systematic uncertainties (e.g due to the uncertainties on the source size). The combined object size for each species is indicated by the dashed lines and the bands represents the corresponding uncertainties.}
  \label{fig:invariant_mass}
\end{figure}

Equation~\ref{coal2_Hyp}  is fitted to the measured  particle ratios with $\rrhyp$ as a free parameter resulting in an average separation between the deuteron and the $\Lambda$ of $9.54^{+0.88}_{-0.70}$ fm. The given uncertainties originate from the fit and take into account all statistical and systematic uncertainties discussed above.
Additional systematic uncertainties, related to the source size, the coalescence model used, and the usage of a specific wave function, are ${}^{+2.52}_{-0.86}$ fm. A detailed description of the procedure can be found in the Methods section~\ref{methods}. 
Similar expressions exist in the coalescence model to calculate the d/p and \he/p ratios~~\cite{Sun:2018mqq} 
which are used to validate the procedure. 
The radii of deuteron and \he\ are determined by the same method as described above, using previous ALICE measurements of the d/p~\cite{ALICE:2019bnp,ALICE:2019dgz,ALICE:2020foi,ALICE:2021ovi} and the \he/p~\cite{ALICE:2017xrp,ALICE:2021ovi,ALICE:2021mfm} ratios at different multiplicities, as shown in Fig.~\ref{fig:invariant_mass}.
The details of the fits are described in the Methods part~\ref{methods}.
A matter radius of ${1.99}^{+0.30}_{-0.29}$ fm is obtained for the deuteron, which is in good agreement with the reference value of
1.96~fm~\cite{PhysRevC.79.014002,Carlson:2014vla}.\he\ is known to be a compact object, and it is not assumed to contain a deuteron subsystem. Consequently, the three-body coalescence is used, resulting in a matter radius of ${2.26}^{+0.28}_{-0.27}$ fm, which is also compatible with the reference value of 1.76 fm~\cite{PhysRevC.79.014002,Carlson:2014vla}. The good agreement of the fit results with the reference values confirms the validity of the wave-function femtometry approach.

\begin{figure}[tb]
    \begin{center}
    \includegraphics[width = 0.49\textwidth]{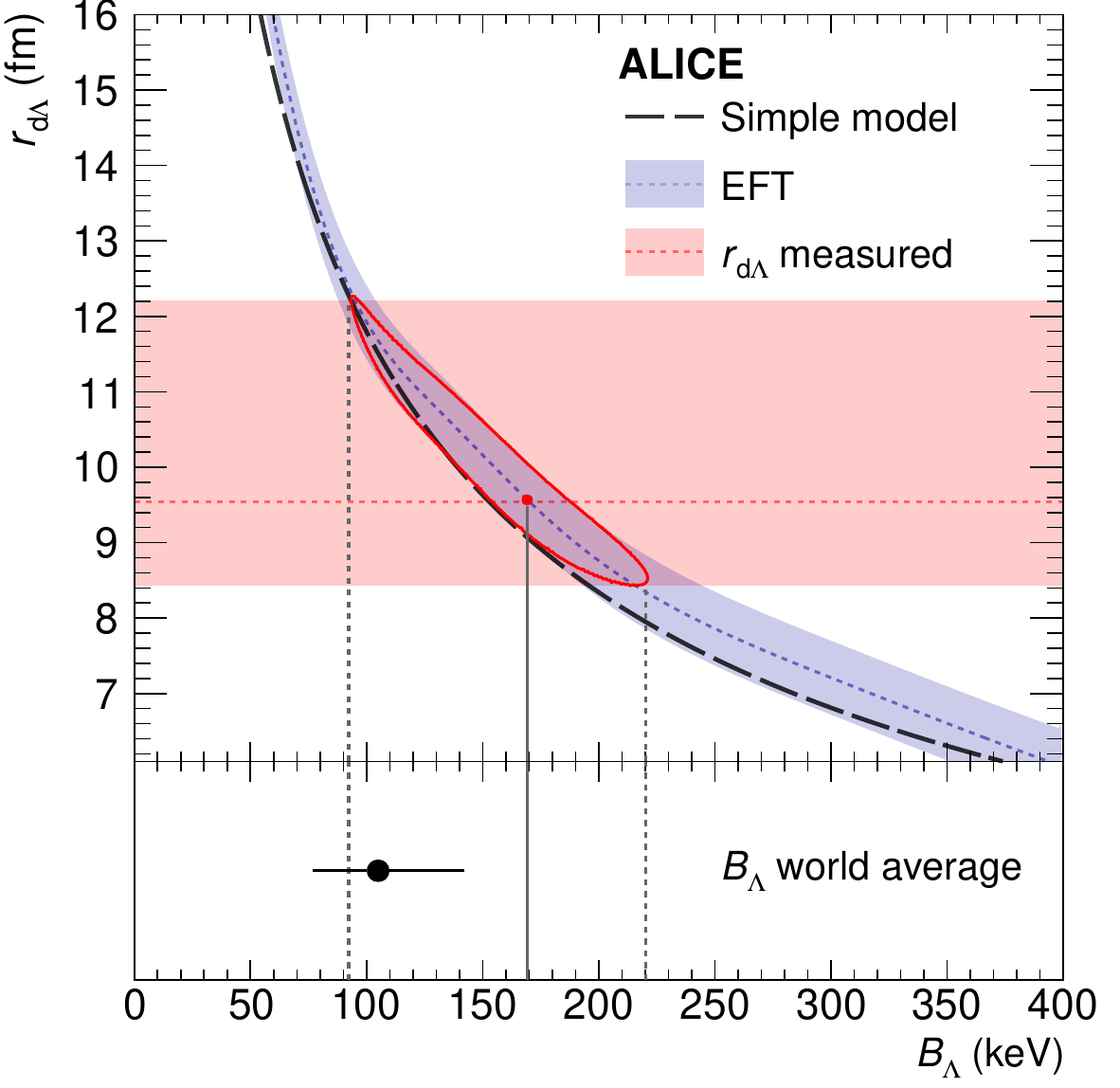}
    \end{center}
    \caption{Hypertriton radius calculated from the $\Lambda$ separation energy using the pionless EFT~\cite{Hildenbrand:2019sgp} (blue band) and a simple quantum mechanical model~\cite{Braun-Munzinger:2018hat} (dashed black line). The measured hypertriton radius is shown as a red band. The red point represents $B_\Lambda$, which is determined  from the intersection of the central values of the EFT calculation and the measured radius. The red contour is the total uncertainty of the measurement. The vertical lines represent the obtained value of $B_\Lambda$ (middle full line) and its uncertainty (outer dashed lines). In the lower panel the $B_\Lambda$ world average~\cite{HypernuclearDataBase} is shown.}
    
    \label{fig:B_Lambda}
\end{figure}

The determination of the hypertriton matter radius using wave-function femtometry is the first experimental verification of the halo nature of a hypernucleus. Earlier estimates of the hypertriton radius~\cite{Hildenbrand:2019sgp} are derived from measurements of the $\Lambda$-separation energy ($B_\Lambda$) using pionless effective field theory.
The present direct measurement of the hypertriton radius makes it possible to reverse this calculation and evaluate $B_\Lambda$ from the measured radius.
The  value of $B_\Lambda = 169^{+51}_{-77}$ keV 
obtained from this procedure is in good agreement with the world average of $B_\Lambda = 105^{+37}_{-28}$~keV~\cite{HypernuclearDataBase}, as shown in Figure~\ref{fig:B_Lambda}. 

\section{Summary}
In this work, the first measurement of hypertriton production in pp collisions is presented. It is shown with the greatest significance to date that the production of nuclear clusters follows the characteristic  system size dependence expected in the nuclear coalescence picture. Canonical implementations of statistical hadronization models, on the other hand, cannot describe the data well in the range of small charged-particle multiplicities. The good description within the coalescence approach justifies using the size of the cluster as a free parameter to achieve an optimal fit of the coalescence curve to the data. With this method, which we call wave-function femtometry, the first direct measurement of the size of the hypertriton is possible. The measured distance between deuteron and $\Lambda$ of $9.54^{+2.67}_{-1.11}$ fm confirms the halo nature of the hypertriton. If the wave-function femtometry method is applied in an analogous manner to the measured d/p and \he/p ratios, the literature values for the deuteron and \he~radii are well reproduced within the uncertainties. 
The newly-established wave-function femtometry method opens a new field in the spectroscopy of light (hyper) nuclei and exotic objects. The system size-dependent measurement of production rates will give complementary access to nuclear matter radii and halo properties. With the upgraded ALICE detector and new next-generation experiments, it will be possible to expand studies to $A=4$ hypernuclei and charmed nuclei (nuclei that contain baryons with at least one charm quark)~\cite{Adamova:2019vkf,ALICE:2022wwr,ALICE:2023bsp}. Moreover, the study of exotic objects such as tetraquarks and pentaquarks may shed light on their nature in terms of hadro-molecules or compact multi-quark states. One example here is the $\chi_{c1}(3872)$ (also known as X(3872)), which is expected to be very weakly bound with a spatially-extended wave function in case the molecular assumption is correct~\cite{Artoisenet:2010uu,Bellini:2018epz}.  

%%%%%%%%%%%%%%%%%%%%%%%%%%%%%%%%
% end main text 
%%%%%%%%%%%%%%%%%%%%%%%%%%%%%%%%

%%%%% acknowledgements - handled by EB chairs 
\newenvironment{acknowledgement}{\relax}{\relax}
\begin{acknowledgement}
\section*{Acknowledgements}
The authors thank Kai-Jia Sun, Avraham Gal, Hans-Werner Hammer, and Fabian Hildenbrand for useful correspondence. 
% add specific acknowledgements here 
% ...but please don't remove the line below: funding agencies
% will be acknowledged with a custom tex file handled by EB chairs after Collab Round 2
% Version: 2026-02-13

The ALICE Collaboration would like to thank all its engineers and technicians for their invaluable contributions to the construction of the experiment and the CERN accelerator teams for the outstanding performance of the LHC complex.
The ALICE Collaboration gratefully acknowledges the resources and support provided by all Grid centres and the Worldwide LHC Computing Grid (WLCG) collaboration.
The ALICE Collaboration acknowledges the following funding agencies for their support in building and running the ALICE detector:
A. I. Alikhanyan National Science Laboratory (Yerevan Physics Institute) Foundation (ANSL), State Committee of Science and World Federation of Scientists (WFS), Armenia;
Austrian Academy of Sciences, Austrian Science Fund (FWF): [M 2467-N36] and Nationalstiftung f\"{u}r Forschung, Technologie und Entwicklung, Austria;
Ministry of Communications and High Technologies, National Nuclear Research Center, Azerbaijan;
Rede Nacional de Física de Altas Energias (Renafae), Financiadora de Estudos e Projetos (Finep), Funda\c{c}\~{a}o de Amparo \`{a} Pesquisa do Estado de S\~{a}o Paulo (FAPESP) and The Sao Paulo Research Foundation  (FAPESP), Brazil;
Bulgarian Ministry of Education and Science, within the National Roadmap for Research Infrastructures 2020-2027 (object CERN), Bulgaria;
Ministry of Education of China (MOEC) , Ministry of Science \& Technology of China (MSTC) and National Natural Science Foundation of China (NSFC), China;
Ministry of Science and Education and Croatian Science Foundation, Croatia;
Centro de Aplicaciones Tecnol\'{o}gicas y Desarrollo Nuclear (CEADEN), Cubaenerg\'{\i}a, Cuba;
Ministry of Education, Youth and Sports of the Czech Republic, Czech Republic;
The Danish Council for Independent Research | Natural Sciences, the VILLUM FONDEN and Danish National Research Foundation (DNRF), Denmark;
Helsinki Institute of Physics (HIP), Finland;
Commissariat \`{a} l'Energie Atomique (CEA) and Institut National de Physique Nucl\'{e}aire et de Physique des Particules (IN2P3) and Centre National de la Recherche Scientifique (CNRS), France;
Bundesministerium f\"{u}r Forschung, Technologie und Raumfahrt (BMFTR) and GSI Helmholtzzentrum f\"{u}r Schwerionenforschung GmbH, Germany;
National Research, Development and Innovation Office, Hungary;
Department of Atomic Energy Government of India (DAE), Department of Science and Technology, Government of India (DST), University Grants Commission, Government of India (UGC) and Council of Scientific and Industrial Research (CSIR), India;
National Research and Innovation Agency - BRIN, Indonesia;
Istituto Nazionale di Fisica Nucleare (INFN), Italy;
Japanese Ministry of Education, Culture, Sports, Science and Technology (MEXT) and Japan Society for the Promotion of Science (JSPS) KAKENHI, Japan;
Consejo Nacional de Ciencia (CONACYT) y Tecnolog\'{i}a, through Fondo de Cooperaci\'{o}n Internacional en Ciencia y Tecnolog\'{i}a (FONCICYT) and Direcci\'{o}n General de Asuntos del Personal Academico (DGAPA), Mexico;
Nederlandse Organisatie voor Wetenschappelijk Onderzoek (NWO), Netherlands;
The Research Council of Norway, Norway;
Pontificia Universidad Cat\'{o}lica del Per\'{u}, Peru;
Ministry of Science and Higher Education, National Science Centre and WUT ID-UB, Poland;
Korea Institute of Science and Technology Information and National Research Foundation of Korea (NRF), Republic of Korea;
Ministry of Education and Scientific Research, Institute of Atomic Physics, Ministry of Research and Innovation and Institute of Atomic Physics and Universitatea Nationala de Stiinta si Tehnologie Politehnica Bucuresti, Romania;
Ministerstvo skolstva, vyskumu, vyvoja a mladeze SR, Slovakia;
National Research Foundation of South Africa, South Africa;
Swedish Research Council (VR) and Knut \& Alice Wallenberg Foundation (KAW), Sweden;
European Organization for Nuclear Research, Switzerland;
Suranaree University of Technology (SUT), National Science and Technology Development Agency (NSTDA) and National Science, Research and Innovation Fund (NSRF via PMU-B B05F650021), Thailand;
Turkish Energy, Nuclear and Mineral Research Agency (TENMAK), Turkey;
National Academy of  Sciences of Ukraine, Ukraine;
Science and Technology Facilities Council (STFC), United Kingdom;
National Science Foundation of the United States of America (NSF) and United States Department of Energy, Office of Nuclear Physics (DOE NP), United States of America.
In addition, individual groups or members have received support from:
FORTE project, reg.\ no.\ CZ.02.01.01/00/22\_008/0004632, Czech Republic, co-funded by the European Union, Czech Republic;
European Research Council (grant no. 950692), European Union;
Deutsche Forschungs Gemeinschaft (DFG, German Research Foundation) ``Neutrinos and Dark Matter in Astro- and Particle Physics'' (grant no. SFB 1258), Germany;
FAIR - Future Artificial Intelligence Research, funded by the NextGenerationEU program (Italy).

\end{acknowledgement}

%%%%%%%% Bibliography 
\bibliographystyle{utphys}   % Remember we use title in the biblio
\bibliography{bibliography}

\providecommand{\href}[2]{#2}\begingroup\raggedright\begin{thebibliography}{100}

\bibitem{Krane:1987ky}
K.~S. Krane, {\em {INTRODUCTORY NUCLEAR PHYSICS}}.
\newblock 1987.

\bibitem{Obertelli:2021crc}
A.~Obertelli and H.~Sagawa,
  \href{https://doi.org/10.1007/978-981-16-2289-2}{{\em {Modern Nuclear
  Physics. From Fundamentals to Frontiers}}}.
\newblock UNITEXT for Physics. Springer, 2021.

\bibitem{Tolos:2020aln}
L.~Tolos and L.~Fabbietti, ``{Strangeness in Nuclei and Neutron Stars}'',
  \href{https://doi.org/10.1016/j.ppnp.2020.103770}{{\em Prog. Part. Nucl.
  Phys.} {\bfseries 112} (2020) 103770},
  \href{https://arxiv.org/abs/2002.09223}{{\ttfamily arXiv:2002.09223
  [nucl-ex]}}.

\bibitem{Burgio:2021vgk}
G.~F. Burgio, H.~J. Schulze, I.~Vidana, and J.~B. Wei, ``{Neutron stars and the
  nuclear equation of state}'',
  \href{https://doi.org/10.1016/j.ppnp.2021.103879}{{\em Prog. Part. Nucl.
  Phys.} {\bfseries 120} (2021) 103879},
  \href{https://arxiv.org/abs/2105.03747}{{\ttfamily arXiv:2105.03747
  [nucl-th]}}.

\bibitem{Vidana:2024ngv}
I.~Vidana, V.~M. Sarti, J.~Haidenbauer, D.~L. Mihaylov, and L.~Fabbietti,
  ``{Neutron Star Properties and Femtoscopic Constraints}'',
  \href{https://doi.org/10.1140/epja/s10050-025-01539-z}{{\em Eur. Phys. J. A}
  {\bfseries 61} (2025) 59}, \href{https://arxiv.org/abs/2412.12729}{{\ttfamily
  arXiv:2412.12729 [nucl-th]}}.

\bibitem{ParticleDataGroup:2022pth}
{\bfseries Particle Data Group} Collaboration, R.~L. Workman {\em et~al.},
  ``{Review of Particle Physics}'',
  \href{https://doi.org/10.1093/ptep/ptac097}{{\em PTEP} {\bfseries 2022}
  (2022) 083C01}.

\bibitem{Haidenbauer:2023qhf}
J.~Haidenbauer, U.-G. Mei\ss{}ner, A.~Nogga, and H.~Le,
  ``{Hyperon\textendash{}nucleon interaction in chiral effective field theory
  at next-to-next-to-leading order}'',
  \href{https://doi.org/10.1140/epja/s10050-023-00960-6}{{\em Eur. Phys. J. A}
  {\bfseries 59} (2023) 63}, \href{https://arxiv.org/abs/2301.00722}{{\ttfamily
  arXiv:2301.00722 [nucl-th]}}.

\bibitem{Mihaylov:2023ahn}
D.~L. Mihaylov, J.~Haidenbauer, and V.~M. Sarti, ``{Constraining the
  p\ensuremath{\Lambda} interaction from a combined analysis of scattering data
  and correlation functions}'',
  \href{https://doi.org/10.1016/j.physletb.2024.138550}{{\em Phys. Lett. B}
  {\bfseries 850} (2024) 138550},
  \href{https://arxiv.org/abs/2312.16970}{{\ttfamily arXiv:2312.16970
  [nucl-th]}}.

\bibitem{ALICE:2019eol}
{\bfseries ALICE} Collaboration, S.~Acharya {\em et~al.}, ``{Study of the
  $\Lambda$--$\Lambda$ interaction with femtoscopy correlations in pp and p--Pb
  collisions at the LHC}'',
  \href{https://doi.org/10.1016/j.physletb.2019.134822}{{\em Phys. Lett. B}
  {\bfseries 797} (2019) 134822},
  \href{https://arxiv.org/abs/1905.07209}{{\ttfamily arXiv:1905.07209
  [nucl-ex]}}.

\bibitem{ALICE:2020mfd}
{\bfseries ALICE} Collaboration, A.~Collaboration {\em et~al.}, ``{Unveiling
  the strong interaction among hadrons at the LHC}'',
  \href{https://doi.org/10.1038/s41586-020-3001-6}{{\em Nature} {\bfseries 588}
  (2020) 232--238}, \href{https://arxiv.org/abs/2005.11495}{{\ttfamily
  arXiv:2005.11495 [nucl-ex]}}. [Erratum: Nature 590, E13 (2021)].

\bibitem{RevMod88:Strangeness}
{A. Gal, E.V. Hungerford and D.J. Millener}, ``{Strangeness in nuclear
  physics}'', {\em Rev. Mod. Phys.} {\bfseries 88} (2016) 035004,
  \href{https://arxiv.org/abs/1605.00557}{{\ttfamily arXiv:1605.00557
  [nucl-th]}}.

\bibitem{HypernuclearDataBase}
P.~Eckert, P.~Achenbach, {\em et~al.}, ``Chart of hypernuclides --
  {H}ypernuclear structure and decay data.'' 2023.
\newblock
  \href{https://hypernuclei.kph.uni-mainz.de}{hypernuclei.kph.uni-mainz.de}.

\bibitem{Hammer:2001ng}
H.~W. Hammer, ``{The Hypertriton in effective field theory}'',
  \href{https://doi.org/10.1016/S0375-9474(02)00621-8}{{\em Nucl. Phys. A}
  {\bfseries 705} (2002) 173--189},
  \href{https://arxiv.org/abs/nucl-th/0110031}{{\ttfamily
  arXiv:nucl-th/0110031}}.

\bibitem{Jensen:2004zz}
A.~S. Jensen, K.~Riisager, D.~V. Fedorov, and E.~Garrido, ``{Structure and
  reactions of quantum halos}'',
  \href{https://doi.org/10.1103/RevModPhys.76.215}{{\em Rev. Mod. Phys.}
  {\bfseries 76} (2004) 215--261}.

\bibitem{Braun-Munzinger:2018hat}
{P. Braun-Munzinger and B. D\"{o}nigus}, ``{Loosely-bound objects produced in
  nuclear collisions at the LHC}'',
  \href{https://doi.org/10.1016/j.nuclphysa.2019.02.006}{{\em Nucl. Phys. A}
  {\bfseries 987} (2019) 144--201},
\href{https://arxiv.org/abs/1809.04681}{{\ttfamily arXiv:1809.04681
  [nucl-ex]}}.
%%CITATION = ARXIV:1809.04681;%%.

\bibitem{Hansen:1987mc}
P.~G. Hansen and B.~Jonson, ``{The Neutron halo of extremely neutron-rich
  nuclei}'', \href{https://doi.org/10.1209/0295-5075/4/4/005}{{\em EPL}
  {\bfseries 4} (1987) 409--414}.

\bibitem{Riisager:1994zz}
K.~Riisager, ``{Nuclear halo states}'',
  \href{https://doi.org/10.1103/RevModPhys.66.1105}{{\em Rev. Mod. Phys.}
  {\bfseries 66} (1994) 1105--1116}.

\bibitem{Tanihata:1995yv}
I.~Tanihata, ``{Neutron halo nuclei}'',
  \href{https://doi.org/10.1088/0954-3899/22/2/004}{{\em J. Phys. G} {\bfseries
  22} (1996) 157--198}.

\bibitem{Tanihata:2020atp}
I.~Tanihata, H.~Toki, and T.~Kajino, eds.,
  \href{https://doi.org/10.1007/978-981-15-8818-1}{{\em {Handbook of Nuclear
  Physics}}}.
\newblock Springer, 2023.

\bibitem{ALICE:2022sco}
{\bfseries ALICE} Collaboration, S.~Acharya {\em et~al.}, ``{Measurement of the
  Lifetime and \ensuremath{\Lambda} Separation Energy of $^3_\Lambda$H }'',
  \href{https://doi.org/10.1103/PhysRevLett.131.102302}{{\em Phys. Rev. Lett.}
  {\bfseries 131} (2023) 102302},
  \href{https://arxiv.org/abs/2209.07360}{{\ttfamily arXiv:2209.07360
  [nucl-ex]}}.

\bibitem{Prem:1964hyp}
R.~J. Prem and P.~H. Steinberg, ``{Lifetimes of Hypernuclei,
  $^{3}_{\Lambda}\mathrm{H},\ ^{4}_{\Lambda}\mathrm{H},\
  ^{5}_{\Lambda}\mathrm{H}$}'',
  \href{https://doi.org/10.1103/PhysRev.136.B1803}{{\em Phys. Rev.} {\bfseries
  136} (1964) B1803--B1806}.

\bibitem{Keyes:1968zz}
G.~Keyes, M.~Derrick, T.~Fields, L.~Hyman, J.~Fetkovich, {\em et~al.}, ``{New
  Measurement of the $^3_\Lambda$H Lifetime}'',
\href{https://doi.org/10.1103/PhysRevLett.20.819}{{\em Phys.Rev.Lett.}
  {\bfseries 20} (1968) 819--821}.
%%CITATION = PRLTA,20,819;%%.

\bibitem{Phillips:1969uy}
R.~Phillips and J.~Schneps, ``{Lifetimes of light hyperfragments. II}'',
\href{https://doi.org/10.1103/PhysRev.180.1307}{{\em Phys.Rev.} {\bfseries 180}
  (1969) 1307--1318}.
%%CITATION = PHRVA,180,1307;%%.

\bibitem{Bohm:1970se}
G.~Bohm, J.~Klabuhn, U.~Krecker, F.~Wysotzki, G.~Coremans, {\em et~al.}, ``{On
  the lifetime of the $^3_\Lambda$H\ hypernucleus}'',
\href{https://doi.org/10.1016/0550-3213(70)90335-4}{{\em Nucl.Phys.} {\bfseries
  B16} (1970) 46--52}.
%%CITATION = NUPHA,B16,46;%%.

\bibitem{Keyes:1970ck}
G.~Keyes, M.~Derrick, T.~Fields, L.~Hyman, J.~Fetkovich, {\em et~al.},
  ``{Properties of $^3_\Lambda$H}'',
\href{https://doi.org/10.1103/PhysRevD.1.66}{{\em Phys.Rev.} {\bfseries D1}
  (1970) 66--77}.
%%CITATION = PHRVA,D1,66;%%.

\bibitem{Keyes:1974ev}
G.~Keyes, J.~Sacton, J.~Wickens, and M.~Block, ``{A measurement of the lifetime
  of the $^3_\Lambda$H hypernucleus}'',
\href{https://doi.org/10.1016/0550-3213(73)90197-1}{{\em Nucl.Phys.} {\bfseries
  B67} (1973) 269--283}.
%%CITATION = NUPHA,B67,269;%%.

\bibitem{STAR:2010gyg}
{\bfseries STAR} Collaboration, B.~I. Abelev {\em et~al.}, ``{Observation of an
  Antimatter Hypernucleus}'',
  \href{https://doi.org/10.1126/science.1183980}{{\em Science} {\bfseries 328}
  (2010) 58--62}, \href{https://arxiv.org/abs/1003.2030}{{\ttfamily
  arXiv:1003.2030 [nucl-ex]}}.

\bibitem{Rappold:2013fic}
C.~Rappold, E.~Kim, D.~Nakajima, T.~Saito, O.~Bertini, {\em et~al.},
  ``{Hypernuclear spectroscopy of products from $^{6}$Li projectiles on a
  carbon target at 2 AGeV}'',
  \href{https://doi.org/10.1016/j.nuclphysa.2013.05.019}{{\em Nucl.Phys.}
  {\bfseries A913} (2013) 170--184},
\href{https://arxiv.org/abs/1305.4871}{{\ttfamily arXiv:1305.4871 [nucl-ex]}}.
%%CITATION = ARXIV:1305.4871;%%.

\bibitem{ALICE:2015oer}
{\bfseries ALICE} Collaboration, J.~Adam {\em et~al.},
  ``{$^{3}_{\Lambda}\mathrm H$ and $^{3}_{\bar{\Lambda}} \overline{\mathrm H}$
  production in Pb--Pb collisions at $\sqrt{s_{\rm NN}} =$ 2.76 TeV}'',
  \href{https://doi.org/10.1016/j.physletb.2016.01.040}{{\em Phys. Lett. B}
  {\bfseries 754} (2016) 360--372},
  \href{https://arxiv.org/abs/1506.08453}{{\ttfamily arXiv:1506.08453
  [nucl-ex]}}.

\bibitem{Adamczyk:2017buv}
{\bfseries STAR} Collaboration, L.~Adamczyk {\em et~al.}, ``{Measurement of the
  $^{3}_{\Lambda}\mathrm{H}$ lifetime in Au + Au collisions at the Relativistic
  Heavy-Ion Collider}'', \href{https://doi.org/10.1103/PhysRevC.97.054909}{{\em
  Phys. Rev.} {\bfseries C97} (2018) 054909},
\href{https://arxiv.org/abs/1710.00436}{{\ttfamily arXiv:1710.00436
  [nucl-ex]}}.
%%CITATION = ARXIV:1710.00436;%%.

\bibitem{ALICE:2019vlx}
{\bfseries ALICE} Collaboration, S.~Acharya {\em et~al.},
  ``{$^3_\Lambda\mathrm{H}$ and $^3_{\bar{\Lambda}}\mathrm{\overline{H}}$
  lifetime measurement in Pb--Pb collisions at $\sqrt{s_{\mathrm{NN}}} = $ 5.02
  TeV via two-body decay}'',
  \href{https://doi.org/10.1016/j.physletb.2019.134905}{{\em Phys. Lett. B}
  {\bfseries 797} (2019) 134905},
  \href{https://arxiv.org/abs/1907.06906}{{\ttfamily arXiv:1907.06906
  [nucl-ex]}}.

\bibitem{Cobis:1996ru}
A.~Cobis, A.~S. Jensen, and D.~V. Fedorov, ``{Three body halos. 5. The
  Structure of the hypertriton}'',
  \href{https://doi.org/10.1088/0954-3899/23/4/002}{{\em J. Phys. G} {\bfseries
  23} (1997) 401--421}, \href{https://arxiv.org/abs/nucl-th/9608026}{{\ttfamily
  arXiv:nucl-th/9608026}}.

\bibitem{Nemura:1999qp}
H.~Nemura, Y.~Suzuki, Y.~Fujiwara, and C.~Nakamoto, ``{Study of light Lambda
  and Lambda-Lambda hypernuclei with the stochastic variational method and
  effective Lambda N potentials}'',
  \href{https://doi.org/10.1143/PTP.103.929}{{\em Prog. Theor. Phys.}
  {\bfseries 103} (2000) 929--958},
\href{https://arxiv.org/abs/nucl-th/9912065}{{\ttfamily arXiv:nucl-th/9912065
  [nucl-th]}}.
%%CITATION = NUCL-TH/9912065;%%.

\bibitem{Hildenbrand:2019sgp}
F.~Hildenbrand and H.~W. Hammer, ``{Three-Body Hypernuclei in Pionless
  Effective Field Theory}'',
  \href{https://doi.org/10.1103/PhysRevC.100.034002}{{\em Phys. Rev. C}
  {\bfseries 100} (2019) 034002},
  \href{https://arxiv.org/abs/1904.05818}{{\ttfamily arXiv:1904.05818
  [nucl-th]}}. [Erratum: Phys.Rev.C 102, 039901 (2020)].

\bibitem{Bertulani:2022vad}
C.~A. Bertulani, ``{Probing the size and binding energy of the hypertriton in
  heavy ion collisions}'',
  \href{https://doi.org/10.1016/j.physletb.2022.137639}{{\em Phys. Lett. B}
  {\bfseries 837} (2023) 137639},
  \href{https://arxiv.org/abs/2211.12643}{{\ttfamily arXiv:2211.12643
  [nucl-th]}}.

\bibitem{Skawran:2019krq}
A.~Skawran {\em et~al.}, ``{Towards nuclear structure with radioactive muonic
  atoms}'', \href{https://doi.org/10.1393/ncc/i2019-19125-7}{{\em Nuovo Cim. C}
  {\bfseries 42} (2019) 125}.

\bibitem{Grinin:2020txk}
A.~Grinin, A.~Matveev, D.~C. Yost, L.~Maisenbacher, V.~Wirthl, R.~Pohl, T.~W.
  H\"ansch, and T.~Udem, ``{Two-photon frequency comb spectroscopy of atomic
  hydrogen}'', \href{https://doi.org/10.1126/science.abc7776}{{\em Science}
  {\bfseries 370} (2020) abc7776}.

\bibitem{Krauth:2021foz}
J.~J. Krauth {\em et~al.}, ``{Measuring the \ensuremath{\alpha}-particle charge
  radius with muonic helium-4 ions}'',
  \href{https://doi.org/10.1038/s41586-021-03183-1}{{\em Nature} {\bfseries
  589} (2021) 527--531}.

\bibitem{Velardita:2023lbi}
S.~Velardita, H.~Alvarez-Pol, T.~Aumann, Y.~Ayyad, M.~Duer, H.-W. Hammer,
  L.~Ji, A.~Obertelli, and Y.~Sun, ``{Method to evidence hypernuclear halos
  from a two-target interaction cross section measurement}'',
  \href{https://doi.org/10.1140/epja/s10050-023-01050-3}{{\em Eur. Phys. J. A}
  {\bfseries 59} (2023) 139}.

\bibitem{Juric:1973zq}
M.~Juric {\em et~al.}, ``{A new determination of the binding-energy values of
  the light hypernuclei (\ensuremath{A\leq 15})}'',
  \href{https://doi.org/10.1016/0550-3213(73)90084-9}{{\em Nucl. Phys. B}
  {\bfseries 52} (1973) 1--30}.

\bibitem{Hildenbrand:2023rkc}
F.~Hildenbrand and H.-W. Hammer, ``{Pionic final state interactions and the
  hypertriton lifetime}'',
  \href{https://doi.org/10.1140/epja/s10050-023-01197-z}{{\em Eur. Phys. J. A}
  {\bfseries 59} (2023) 280},
  \href{https://arxiv.org/abs/2309.12822}{{\ttfamily arXiv:2309.12822
  [nucl-th]}}.

\bibitem{Angeli:2013epw}
I.~Angeli and K.~P. Marinova, ``{Table of experimental nuclear ground state
  charge radii: An update}'',
  \href{https://doi.org/10.1016/j.adt.2011.12.006}{{\em Atom. Data Nucl. Data
  Tabl.} {\bfseries 99} (2013) 69--95}.

\bibitem{butler_pearson61}
S.~T. Butler and C.~A. Pearson, ``Deuterons from high-energy proton bombardment
  of matter'', \href{https://doi.org/10.1103/PhysRevLett.7.69}{{\em Phys. Rev.
  Lett.} {\bfseries 7} (Jul, 1961) 69--71}.
  \url{http://link.aps.org/doi/10.1103/PhysRevLett.7.69}.

\bibitem{Butler:1963pp}
S.~Butler and C.~Pearson, ``Deuterons from high-energy proton bombardment of
  matter'',
\href{https://doi.org/10.1103/PhysRev.129.836}{{\em Phys. Rev.} {\bfseries 129}
  (1963) 836--842}.
%%CITATION = PHRVA,129,836;%%.

\bibitem{ALICE:2025zzg}
{\bfseries ALICE} Collaboration, S.~Acharya {\em et~al.}, ``{Accessing the
  deuteron source with pion-deuteron femtoscopy in Pb-Pb collisions at
  $\sqrt{s_{\rm NN}} = 5.02$ TeV}'',
  \href{https://doi.org/10.1038/s41586-025-09775-5}{{\em Nature} {\bfseries
  648} (2025) 306–--311}, \href{https://arxiv.org/abs/2504.02333}{{\ttfamily
  arXiv:2504.02333 [nucl-ex]}}.

\bibitem{Scheibl:1998tk}
R.~Scheibl and U.~W. Heinz, ``{Coalescence and flow in ultrarelativistic heavy
  ion collisions}'', \href{https://doi.org/10.1103/PhysRevC.59.1585}{{\em Phys.
  Rev. C} {\bfseries 59} (1999) 1585--1602},
\href{https://arxiv.org/abs/nucl-th/9809092}{{\ttfamily arXiv:nucl-th/9809092
  [nucl-th]}}.
%%CITATION = NUCL-TH/9809092;%%.

\bibitem{Bellini:2018epz}
F.~Bellini and A.~P. Kalweit, ``{Testing production scenarios for
  (anti-)(hyper-)nuclei and exotica at energies available at the CERN Large
  Hadron Collider}'', \href{https://doi.org/10.1103/PhysRevC.99.054905}{{\em
  Phys. Rev. C} {\bfseries 99} (2019) 054905},
  \href{https://arxiv.org/abs/1807.05894}{{\ttfamily arXiv:1807.05894
  [hep-ph]}}.

\bibitem{Blum:2017iwq}
K.~Blum, R.~Sato, and E.~Waxman, ``{Cosmic-ray Antimatter}'',
  \href{https://arxiv.org/abs/1709.06507}{{\ttfamily arXiv:1709.06507
  [astro-ph.HE]}}.

\bibitem{Bellini:2020cbj}
F.~Bellini, K.~Blum, A.~P. Kalweit, and M.~Puccio, ``{Examination of
  coalescence as the origin of nuclei in hadronic collisions}'',
  \href{https://doi.org/10.1103/PhysRevC.103.014907}{{\em Phys. Rev. C}
  {\bfseries 103} (2021) 014907},
  \href{https://arxiv.org/abs/2007.01750}{{\ttfamily arXiv:2007.01750
  [nucl-th]}}.

\bibitem{Mahlein:2023fmx}
M.~Mahlein, L.~Barioglio, F.~Bellini, L.~Fabbietti, C.~Pinto, B.~Singh, and
  S.~Tripathy, ``{A realistic coalescence model for deuteron production}'',
  \href{https://doi.org/10.1140/epjc/s10052-023-11972-3}{{\em Eur. Phys. J. C}
  {\bfseries 83} (2023) 804},
  \href{https://arxiv.org/abs/2302.12696}{{\ttfamily arXiv:2302.12696
  [hep-ex]}}.

\bibitem{Mahlein:2024pur}
M.~Mahlein, C.~Pinto, and L.~Fabbietti, ``{ToMCCA: A Toy Monte Carlo
  Coalescence Afterburner}'',
  \href{https://arxiv.org/abs/2404.03352}{{\ttfamily arXiv:2404.03352
  [hep-ph]}}.

\bibitem{Sun:2018mqq}
K.-J. Sun, C.~M. Ko, and B.~D\"onigus, ``{Suppression of light nuclei
  production in collisions of small systems at the Large Hadron Collider}'',
  \href{https://doi.org/10.1016/j.physletb.2019.03.033}{{\em Phys. Lett. B}
  {\bfseries 792} (2019) 132--137},
  \href{https://arxiv.org/abs/1812.05175}{{\ttfamily arXiv:1812.05175
  [nucl-th]}}.

\bibitem{Mahlein:2025bla}
M.~Mahlein, B.~Singh, M.~Viviani, F.~Bellini, L.~Fabbietti, A.~Kievsky, and
  L.~E. Marcucci, ``{ToMCCA-3: A realistic 3-body coalescence model}'',
  \href{https://arxiv.org/abs/2504.02491}{{\ttfamily arXiv:2504.02491
  [hep-ph]}}.

\bibitem{Leung:2025jwe}
Y.~H. Leung, Y.~Zhou, and N.~Herrmann, ``{A Data-Guided Coalescence Model for
  Light Nuclei and Hypernuclei: Validation and Predictions}'',
  \href{https://arxiv.org/abs/2510.06758}{{\ttfamily arXiv:2510.06758
  [nucl-th]}}.

\bibitem{ALICE:2021mfm}
{\bfseries ALICE} Collaboration, S.~Acharya {\em et~al.}, ``{Production of
  light (anti)nuclei in pp collisions at $ \sqrt{s} $ = 13 TeV}'',
  \href{https://doi.org/10.1007/JHEP01(2022)106}{{\em JHEP} {\bfseries 01}
  (2022) 106}, \href{https://arxiv.org/abs/2109.13026}{{\ttfamily
  arXiv:2109.13026 [nucl-ex]}}.

\bibitem{Andronic:2010qu}
A.~Andronic, P.~Braun-Munzinger, J.~Stachel, and H.~Stocker, ``{Production of
  light nuclei, hypernuclei and their antiparticles in relativistic nuclear
  collisions}'', \href{https://doi.org/10.1016/j.physletb.2011.01.053}{{\em
  Phys. Lett. B} {\bfseries 697} (2011) 203--207},
\href{https://arxiv.org/abs/1010.2995}{{\ttfamily arXiv:1010.2995 [nucl-th]}}.
%%CITATION = ARXIV:1010.2995;%%.

\bibitem{Andronic:2017pug}
A.~Andronic, P.~Braun-Munzinger, K.~Redlich, and J.~Stachel, ``{Decoding the
  phase structure of QCD via particle production at high energy}'',
  \href{https://doi.org/10.1038/s41586-018-0491-6}{{\em Nature} {\bfseries 561}
  (2018) 321--330}, \href{https://arxiv.org/abs/1710.09425}{{\ttfamily
  arXiv:1710.09425 [nucl-th]}}.

\bibitem{Donigus:2020ctf}
B.~D\"onigus, ``{Light nuclei in the hadron resonance gas}'',
  \href{https://doi.org/10.1142/S0218301320400017}{{\em Int. J. Mod. Phys. E}
  {\bfseries 29} (2020) 2040001},
  \href{https://arxiv.org/abs/2004.10544}{{\ttfamily arXiv:2004.10544
  [nucl-th]}}.

\bibitem{Donigus:2022haq}
B.~D\"onigus, G.~R\"opke, and D.~Blaschke, ``{Deuteron yields from heavy-ion
  collisions at energies available at the CERN Large Hadron Collider: Continuum
  correlations and in-medium effects}'',
  \href{https://doi.org/10.1103/PhysRevC.106.044908}{{\em Phys. Rev. C}
  {\bfseries 106} (2022) 044908},
  \href{https://arxiv.org/abs/2206.10376}{{\ttfamily arXiv:2206.10376
  [nucl-th]}}.

\bibitem{ALICE:2024koa}
{\bfseries ALICE} Collaboration, S.~Acharya {\em et~al.}, ``{Measurement of
  ${}_{\Lambda}^{3}\mathrm{H}$ production in Pb--Pb collisions at
  $\sqrt{s_{\mathrm{NN}}}$ = 5.02 TeV}'',
  \href{https://doi.org/https://doi.org/10.1016/j.physletb.2024.139066}{{\em
  Physics Letters B} {\bfseries 860} (2025) 139066}.

\bibitem{Vovchenko:2018fiy}
V.~Vovchenko, B.~D\"onigus, and H.~Stoecker, ``{Multiplicity dependence of
  light nuclei production at LHC energies in the canonical statistical
  model}'', \href{https://doi.org/10.1016/j.physletb.2018.08.041}{{\em Phys.
  Lett. B} {\bfseries 785} (2018) 171--174},
  \href{https://arxiv.org/abs/1808.05245}{{\ttfamily arXiv:1808.05245
  [hep-ph]}}.

\bibitem{Cleymans:2020fsc}
J.~Cleymans, P.~M. Lo, K.~Redlich, and N.~Sharma, ``{Multiplicity dependence of
  (multi)strange baryons in the canonical ensemble with phase shift
  corrections}'', \href{https://doi.org/10.1103/PhysRevC.103.014904}{{\em Phys.
  Rev. C} {\bfseries 103} (2021) 014904},
  \href{https://arxiv.org/abs/2009.04844}{{\ttfamily arXiv:2009.04844
  [hep-ph]}}.

\bibitem{Sharma:2022poi}
N.~Sharma, L.~Kumar, P.~M. Lo, and K.~Redlich, ``{Light-nuclei production in pp
  and pA collisions in the baryon canonical ensemble approach}'',
  \href{https://doi.org/10.1103/PhysRevC.107.054903}{{\em Phys. Rev. C}
  {\bfseries 107} (2023) 054903},
  \href{https://arxiv.org/abs/2210.15617}{{\ttfamily arXiv:2210.15617
  [nucl-th]}}.

\bibitem{ALargeIonColliderExperiment:2021puh}
{\bfseries ALICE} Collaboration, S.~Acharya {\em et~al.}, ``{Hypertriton
  Production in p--Pb Collisions at $\sqrt {s_{\mathrm{NN}}}$=5.02\,\,TeV}'',
  \href{https://doi.org/10.1103/PhysRevLett.128.252003}{{\em Phys. Rev. Lett.}
  {\bfseries 128} (2022) 252003},
  \href{https://arxiv.org/abs/2107.10627}{{\ttfamily arXiv:2107.10627
  [nucl-ex]}}.

\bibitem{Aamodt:2008zz}
{\bfseries ALICE} Collaboration, K.~Aamodt {\em et~al.}, ``{The ALICE
  experiment at the CERN LHC}'',
\href{https://doi.org/10.1088/1748-0221/3/08/S08002}{{\em JINST} {\bfseries 3}
  (2008) S08002}.
%%CITATION = JINST,3,S08002;%%.

\bibitem{Abelev:2014ffa}
{\bfseries ALICE} Collaboration, B.~B. Abelev {\em et~al.}, ``{Performance of
  the ALICE Experiment at the CERN LHC}'',
  \href{https://doi.org/10.1142/S0217751X14300440}{{\em Int. J. Mod. Phys. A}
  {\bfseries 29} (2014) 1430044},
\href{https://arxiv.org/abs/1402.4476}{{\ttfamily arXiv:1402.4476 [nucl-ex]}}.
%%CITATION = ARXIV:1402.4476;%%.

\bibitem{ALICE:2022veq}
{\bfseries ALICE} Collaboration, S.~Acharya {\em et~al.}, ``{Light (anti)nuclei
  production in Pb--Pb collisions at $\sqrt{s_\mathrm{NN}}=5.02$~TeV}'',
  \href{https://doi.org/10.1103/PhysRevC.107.064904}{{\em Phys. Rev. C}
  {\bfseries 107} (2023) 064904},
  \href{https://arxiv.org/abs/2211.14015}{{\ttfamily arXiv:2211.14015
  [nucl-ex]}}.

\bibitem{ALICE:2023zbh}
{\bfseries ALICE} Collaboration, S.~Acharya {\em et~al.}, ``{Femtoscopic
  correlations of identical charged pions and kaons in pp collisions at
  $\sqrt{s}=13$ TeV with event-shape selection}'',
  \href{https://doi.org/10.1103/PhysRevC.109.024915}{{\em Phys. Rev. C}
  {\bfseries 109} (2024) 024915},
  \href{https://arxiv.org/abs/2310.07509}{{\ttfamily arXiv:2310.07509
  [nucl-ex]}}.

\bibitem{ALICE:2023sjd}
{\bfseries ALICE} Collaboration, S.~Acharya {\em et~al.}, ``{Common femtoscopic
  hadron-emission source in pp collisions at the LHC}'',
  \href{https://doi.org/10.1140/epjc/s10052-025-13793-y}{{\em Eur. Phys. J. C}
  {\bfseries 85} (2025) 198},
  \href{https://arxiv.org/abs/2311.14527}{{\ttfamily arXiv:2311.14527
  [hep-ph]}}.

\bibitem{Blum:2017qnn}
K.~Blum, K.~C.~Y. Ng, R.~Sato, and M.~Takimoto, ``{Cosmic rays, antihelium, and
  an old navy spotlight}'', {\em Phys. Rev. D} {\bfseries 96} (2017) 103021,
\href{https://arxiv.org/abs/1704.05431}{{\ttfamily arXiv:1704.05431
  [astro-ph.HE]}}.
%%CITATION = ARXIV:1704.05431;%%.

\bibitem{Kamada1595}
H.~Kamada, J.~Golak, K.~Miyagawa, H.~Wita\l{}a, and W.~Gl\"ockle,
  ``\ensuremath{\pi}-mesonic decay of the hypertriton'',
  \href{https://doi.org/10.1103/PhysRevC.57.1595}{{\em Phys. Rev. C} {\bfseries
  57} (Apr, 1998) 1595--1603}.
  \url{https://link.aps.org/doi/10.1103/PhysRevC.57.1595}.

\bibitem{Cranmer:2000du}
K.~S. Cranmer, ``{Kernel estimation in high-energy physics}'',
  \href{https://doi.org/10.1016/S0010-4655(00)00243-5}{{\em Comput. Phys.
  Commun.} {\bfseries 136} (2001) 198--207},
  \href{https://arxiv.org/abs/hep-ex/0011057}{{\ttfamily
  arXiv:hep-ex/0011057}}.

\bibitem{Verkerke:2003ir}
W.~Verkerke and D.~P. Kirkby, ``{The RooFit toolkit for data modeling}'', {\em
  eConf} {\bfseries C0303241} (2003) MOLT007,
  \href{https://arxiv.org/abs/physics/0306116}{{\ttfamily
  arXiv:physics/0306116}}.

\bibitem{Cowan:2010js}
G.~Cowan, K.~Cranmer, E.~Gross, and O.~Vitells, ``{Asymptotic formulae for
  likelihood-based tests of new physics}'',
  \href{https://doi.org/10.1140/epjc/s10052-011-1554-0}{{\em Eur. Phys. J. C}
  {\bfseries 71} (2011) 1554},
  \href{https://arxiv.org/abs/1007.1727}{{\ttfamily arXiv:1007.1727
  [physics.data-an]}}. [Erratum: Eur.Phys.J.C 73, 2501 (2013)].

\bibitem{Vovchenko:2019kes}
{V. Vovchenko, B. D\"{o}nigus and H. Stoecker}, ``{Canonical statistical model
  analysis of p--p, p--Pb, and Pb--Pb collisions at energies available at the
  CERN Large Hadron Collider}'',
  \href{https://doi.org/10.1103/PhysRevC.100.054906}{{\em Phys. Rev. C}
  {\bfseries 100} (2019) 054906},
\href{https://arxiv.org/abs/1906.03145}{{\ttfamily arXiv:1906.03145 [hep-ph]}}.
%%CITATION = ARXIV:1906.03145;%%.

\bibitem{ALICE:2019avo}
{\bfseries ALICE} Collaboration, S.~Acharya {\em et~al.}, ``{Multiplicity
  dependence of (multi-)strange hadron production in proton-proton collisions
  at $\sqrt{s}$ = 13 TeV}'',
  \href{https://doi.org/10.1140/epjc/s10052-020-7673-8}{{\em Eur. Phys. J. C}
  {\bfseries 80} (2020) 167},
  \href{https://arxiv.org/abs/1908.01861}{{\ttfamily arXiv:1908.01861
  [nucl-ex]}}.

\bibitem{ALICE:2020nkc}
{\bfseries ALICE} Collaboration, S.~Acharya {\em et~al.}, ``{Multiplicity
  dependence of $\pi $, K, and p production in pp collisions at $\sqrt{s} = 13$
  TeV}'', \href{https://doi.org/10.1140/epjc/s10052-020-8125-1}{{\em Eur. Phys.
  J. C} {\bfseries 80} (2020) 693},
  \href{https://arxiv.org/abs/2003.02394}{{\ttfamily arXiv:2003.02394
  [nucl-ex]}}.

\bibitem{ALICE:2020jsh}
{\bfseries ALICE} Collaboration, S.~Acharya {\em et~al.}, ``{Production of
  light-flavor hadrons in pp collisions at $\sqrt{s}~=~7\text { and }\sqrt{s} =
  13 \, \text { TeV} $}'',
  \href{https://doi.org/10.1140/epjc/s10052-020-08690-5}{{\em Eur. Phys. J. C}
  {\bfseries 81} (2021) 256},
  \href{https://arxiv.org/abs/2005.11120}{{\ttfamily arXiv:2005.11120
  [nucl-ex]}}.

\bibitem{Castorina:2013mba}
P.~Castorina and H.~Satz, ``{Causality Constraints on Hadron Production In High
  Energy Collisions}'', \href{https://doi.org/10.1142/S0218301314500190}{{\em
  Int. J. Mod. Phys. E} {\bfseries 23} (2014) 1450019},
  \href{https://arxiv.org/abs/1310.6932}{{\ttfamily arXiv:1310.6932 [hep-ph]}}.

\bibitem{ALICE:2024rnr}
{\bfseries ALICE} Collaboration, S.~Acharya {\em et~al.}, ``{Probing
  Strangeness Hadronization with Event-by-Event Production of Multistrange
  Hadrons}'', \href{https://doi.org/10.1103/PhysRevLett.134.022303}{{\em Phys.
  Rev. Lett.} {\bfseries 134} (2025) 022303},
  \href{https://arxiv.org/abs/2405.19890}{{\ttfamily arXiv:2405.19890
  [nucl-ex]}}.

\bibitem{Reichert:2022mek}
T.~Reichert, J.~Steinheimer, V.~Vovchenko, B.~D\"onigus, and M.~Bleicher,
  ``{Energy dependence of light hypernuclei production in heavy-ion collisions
  from a coalescence and statistical-thermal model perspective}'',
  \href{https://doi.org/10.1103/PhysRevC.107.014912}{{\em Phys. Rev. C}
  {\bfseries 107} (2023) 014912},
  \href{https://arxiv.org/abs/2210.11876}{{\ttfamily arXiv:2210.11876
  [nucl-th]}}.

\bibitem{ALICE:2023qyl}
{\bfseries ALICE} Collaboration, S.~Acharya {\em et~al.}, ``{Measurement of
  (anti)alpha production in central Pb{\textendash}Pb collisions at
  s$\sqrt{s_\mathrm{NN}}=5.02$~ TeV}'',
  \href{https://doi.org/10.1016/j.physletb.2024.138943}{{\em Phys. Lett. B}
  {\bfseries 858} (2024) 138943},
  \href{https://arxiv.org/abs/2311.11758}{{\ttfamily arXiv:2311.11758
  [nucl-ex]}}.

\bibitem{ALICE:2024say}
{\bfseries ALICE} Collaboration, S.~Acharya {\em et~al.}, ``{Measurement of the
  production and elliptic flow of (anti)nuclei in Xe--Xe collisions at
  $\sqrt{s_\mathrm{NN}}=5.44$~TeV}'',
  \href{https://doi.org/10.1103/PhysRevC.110.064901}{{\em Phys. Rev. C}
  {\bfseries 110} (2024) 064901},
  \href{https://arxiv.org/abs/2405.19826}{{\ttfamily arXiv:2405.19826
  [nucl-ex]}}.

\bibitem{ALICE:2019bnp}
{\bfseries ALICE} Collaboration, S.~Acharya {\em et~al.}, ``{Multiplicity
  dependence of light (anti-)nuclei production in p-Pb collisions at
  $\sqrt{s_{\mathrm{NN}}}$ = 5.02 TeV}'',
  \href{https://doi.org/10.1016/j.physletb.2019.135043}{{\em Phys. Lett. B}
  {\bfseries 800} (2020) 135043},
  \href{https://arxiv.org/abs/1906.03136}{{\ttfamily arXiv:1906.03136
  [nucl-ex]}}.

\bibitem{ALICE:2019dgz}
{\bfseries ALICE} Collaboration, S.~Acharya {\em et~al.}, ``{Multiplicity
  dependence of (anti-)deuteron production in pp collisions at $\sqrt{s}$ = 7
  TeV}'', \href{https://doi.org/10.1016/j.physletb.2019.05.028}{{\em Phys.
  Lett. B} {\bfseries 794} (2019) 50--63},
  \href{https://arxiv.org/abs/1902.09290}{{\ttfamily arXiv:1902.09290
  [nucl-ex]}}.

\bibitem{ALICE:2020foi}
{\bfseries ALICE} Collaboration, S.~Acharya {\em et~al.}, ``{(Anti-)deuteron
  production in pp collisions at $\sqrt{s}=13 \ \text {TeV}$}'',
  \href{https://doi.org/10.1140/epjc/s10052-020-8256-4}{{\em Eur. Phys. J. C}
  {\bfseries 80} (2020) 889},
  \href{https://arxiv.org/abs/2003.03184}{{\ttfamily arXiv:2003.03184
  [nucl-ex]}}.

\bibitem{ALICE:2021ovi}
{\bfseries ALICE} Collaboration, S.~Acharya {\em et~al.}, ``{Production of
  light (anti)nuclei in pp collisions at $\sqrt{s} = 5.02$~TeV}'',
  \href{https://doi.org/10.1140/epjc/s10052-022-10241-z}{{\em Eur. Phys. J. C}
  {\bfseries 82} (2022) 289},
  \href{https://arxiv.org/abs/2112.00610}{{\ttfamily arXiv:2112.00610
  [nucl-ex]}}.

\bibitem{ALICE:2017xrp}
{\bfseries ALICE} Collaboration, S.~Acharya {\em et~al.}, ``{Production of
  deuterons, tritons, $^{3}$He nuclei and their antinuclei in pp collisions at
  $\mathbf{\sqrt{{\textit s}}}$ = 0.9, 2.76 and 7 TeV}'',
  \href{https://doi.org/10.1103/PhysRevC.97.024615}{{\em Phys. Rev. C}
  {\bfseries 97} (2018) 024615},
  \href{https://arxiv.org/abs/1709.08522}{{\ttfamily arXiv:1709.08522
  [nucl-ex]}}.

\bibitem{PhysRevC.79.014002}
G.~R\"opke, ``Light nuclei quasiparticle energy shifts in hot and dense nuclear
  matter'', \href{https://doi.org/10.1103/PhysRevC.79.014002}{{\em Phys. Rev.
  C} {\bfseries 79} (Jan, 2009) 014002}.
  \url{https://link.aps.org/doi/10.1103/PhysRevC.79.014002}.

\bibitem{Carlson:2014vla}
J.~Carlson, S.~Gandolfi, F.~Pederiva, S.~C. Pieper, R.~Schiavilla, K.~E.
  Schmidt, and R.~B. Wiringa, ``{Quantum Monte Carlo methods for nuclear
  physics}'', \href{https://doi.org/10.1103/RevModPhys.87.1067}{{\em Rev. Mod.
  Phys.} {\bfseries 87} (2015) 1067},
  \href{https://arxiv.org/abs/1412.3081}{{\ttfamily arXiv:1412.3081
  [nucl-th]}}.

\bibitem{Adamova:2019vkf}
D.~Adamov\'a {\em et~al.}, ``{A next-generation LHC heavy-ion experiment}'',
  \href{https://arxiv.org/abs/1902.01211}{{\ttfamily arXiv:1902.01211
  [physics.ins-det]}}.

\bibitem{ALICE:2022wwr}
{\bfseries ALICE} Collaboration, ``{Letter of intent for ALICE 3: A
  next-generation heavy-ion experiment at the LHC}'',
  \href{https://arxiv.org/abs/2211.02491}{{\ttfamily arXiv:2211.02491
  [physics.ins-det]}}.

\bibitem{ALICE:2023bsp}
{\bfseries ALICE} Collaboration, ``{Upgrade of the ALICE Inner Tracking System
  during LS3: study of physics performance}'',.
  \url{https://cds.cern.ch/record/2868015}.

\bibitem{Artoisenet:2010uu}
P.~Artoisenet and E.~Braaten, ``{Estimating the Production Rate of
  Loosely-bound Hadronic Molecules using Event Generators}'',
  \href{https://doi.org/10.1103/PhysRevD.83.014019}{{\em Phys. Rev. D}
  {\bfseries 83} (2011) 014019},
  \href{https://arxiv.org/abs/1007.2868}{{\ttfamily arXiv:1007.2868 [hep-ph]}}.

\bibitem{ALICE:2017ymw}
{\bfseries ALICE} Collaboration, S.~Acharya {\em et~al.}, ``{The ALICE
  Transition Radiation Detector: construction, operation, and performance}'',
  \href{https://doi.org/10.1016/j.nima.2017.09.028}{{\em Nucl. Instrum. Meth.
  A} {\bfseries 881} (2018) 88--127},
  \href{https://arxiv.org/abs/1709.02743}{{\ttfamily arXiv:1709.02743
  [physics.ins-det]}}.

\bibitem{Abbas:2013taa}
{\bfseries ALICE} Collaboration, E.~Abbas {\em et~al.}, ``{Performance of the
  ALICE VZERO system}'',
  \href{https://doi.org/10.1088/1748-0221/8/10/P10016}{{\em JINST} {\bfseries
  8} (2013) P10016},
\href{https://arxiv.org/abs/1306.3130}{{\ttfamily arXiv:1306.3130 [nucl-ex]}}.
%%CITATION = ARXIV:1306.3130;%%.

\bibitem{Kamada:1997rv}
H.~Kamada, J.~Golak, K.~Miyagawa, H.~Witala, and W.~Gloeckle, ``{Pi mesonic
  decay of the hypertriton}'',
  \href{https://doi.org/10.1103/PhysRevC.57.1595}{{\em Phys. Rev.} {\bfseries
  C57} (1998) 1595--1603},
\href{https://arxiv.org/abs/nucl-th/9709035}{{\ttfamily arXiv:nucl-th/9709035
  [nucl-th]}}.
%%CITATION = NUCL-TH/9709035;%%.

\bibitem{STAR:2017gxa}
{\bfseries STAR} Collaboration, L.~Adamczyk {\em et~al.}, ``{Measurement of the
  $^3_{\Lambda}$H lifetime in Au+Au collisions at the BNL Relativistic Heavy
  Ion Collider}'', \href{https://doi.org/10.1103/PhysRevC.97.054909}{{\em Phys.
  Rev. C} {\bfseries 97} (2018) 054909},
  \href{https://arxiv.org/abs/1710.00436}{{\ttfamily arXiv:1710.00436
  [nucl-ex]}}.

\bibitem{ALICE:2015tra}
{\bfseries ALICE} Collaboration, J.~Adam {\em et~al.}, ``{Centrality dependence
  of pion freeze-out radii in Pb--Pb collisions at $\sqrt{s_{NN}}=$ 2.76
  TeV}'', \href{https://doi.org/10.1103/PhysRevC.93.024905}{{\em Phys. Rev. C}
  {\bfseries 93} (2016) 024905},
  \href{https://arxiv.org/abs/1507.06842}{{\ttfamily arXiv:1507.06842
  [nucl-ex]}}.

\bibitem{Abelev:2013vea}
{\bfseries ALICE} Collaboration, B.~Abelev {\em et~al.}, ``{Centrality
  dependence of $\pi$, K, p production in Pb--Pb collisions at $\sqrt{s_{NN}}$
  = 2.76 TeV}'', \href{https://doi.org/10.1103/PhysRevC.88.044910}{{\em Phys.
  Rev.} {\bfseries C88} (2013) 044910},
\href{https://arxiv.org/abs/1303.0737}{{\ttfamily arXiv:1303.0737 [hep-ex]}}.
%%CITATION = ARXIV:1303.0737;%%.

\bibitem{Adam:2015qaa}
{\bfseries ALICE} Collaboration, J.~Adam {\em et~al.}, ``{Measurement of pion,
  kaon and proton production in proton–proton collisions at $\sqrt{s} = 7$
  TeV}'', \href{https://doi.org/10.1140/epjc/s10052-015-3422-9}{{\em Eur. Phys.
  J.} {\bfseries C75} (2015) 226},
\href{https://arxiv.org/abs/1504.00024}{{\ttfamily arXiv:1504.00024
  [nucl-ex]}}.
%%CITATION = ARXIV:1504.00024;%%.

\bibitem{ALICE:2020ibs}
{\bfseries ALICE} Collaboration, S.~Acharya {\em et~al.}, ``{Search for a
  common baryon source in high-multiplicity pp collisions at the LHC}'',
  \href{https://doi.org/10.1016/j.physletb.2020.135849}{{\em Phys. Lett. B}
  {\bfseries 811} (2020) 135849},
  \href{https://arxiv.org/abs/2004.08018}{{\ttfamily arXiv:2004.08018
  [nucl-ex]}}. [Erratum: Phys.Lett.B 861, 139233 (2025)].

\bibitem{ALICE:2025aur}
{\bfseries ALICE} Collaboration, S.~Acharya {\em et~al.}, ``{Investigating the
  $\mathbf {\text {p--}\pi ^{\pm }}$ and $\mathbf {\text {p--p--}\pi ^{\pm }}$
  dynamics with femtoscopy in pp collisions at $
  {\sqrt{\textit{s}}=13}$~TeV}'',
  \href{https://doi.org/10.1140/epja/s10050-025-01615-4}{{\em Eur. Phys. J. A}
  {\bfseries 61} (2025) 194},
  \href{https://arxiv.org/abs/2502.20200}{{\ttfamily arXiv:2502.20200
  [nucl-ex]}}.

\bibitem{ALICE:2015hvw}
{\bfseries ALICE} Collaboration, J.~Adam {\em et~al.}, ``{One-dimensional pion,
  kaon, and proton femtoscopy in Pb--Pb collisions at $\sqrt{s_{\rm {NN}}}$
  =2.76 TeV}'', \href{https://doi.org/10.1103/PhysRevC.92.054908}{{\em Phys.
  Rev. C} {\bfseries 92} (2015) 054908},
  \href{https://arxiv.org/abs/1506.07884}{{\ttfamily arXiv:1506.07884
  [nucl-ex]}}.

\bibitem{ALICE:2025wuy}
{\bfseries ALICE} Collaboration, S.~Acharya {\em et~al.}, ``{Femtoscopic study
  of the proton-proton and proton-deuteron systems in heavy-ion collisions at
  the LHC}'', \href{https://doi.org/10.1016/j.physletb.2025.139921}{{\em Phys.
  Lett. B} {\bfseries 871} (2025) 139921},
  \href{https://arxiv.org/abs/2505.01061}{{\ttfamily arXiv:2505.01061
  [nucl-ex]}}.

\bibitem{Congleton:1992kk}
J.~Congleton, ``{A Simple model of the hypertriton}'',
\href{https://doi.org/10.1088/0954-3899/18/2/015}{{\em J.Phys.} {\bfseries G18}
  (1992) 339--357}.
%%CITATION = JPAGA,G18,339;%%.

\end{thebibliography}\endgroup
%\input {bibliography.tex}  

%%%%%%%%%%%%%%%%%%%%%%%%%%%%%%%%
% Appendices: yours (if any) + authorlist
%%%%%%%%%%%%%%%%%%%%%%%%%%%%%%%%
\newpage
\appendix %Methods section

\section{Methods}
\label{methods}

\subsection*{Data samples and event selection}
The measurement is based on two data samples of pp collisions recorded with different trigger conditions. The high-multiplicity (HM) data sample has a mean charged-particle multiplicity \avdndeta = 30.8 and the
data sample inspected by the Transition Radiation Detector (TRD)~\cite{ALICE:2017ymw} nuclei trigger (HNU) has a 
mean charged-particle multiplicity 
\avdndeta = 6.9, corresponding to a minimum-bias (MB) multiplicity selection. The given \avdndeta values are corrected for inelastic collisions with at least one charged particle.

Two different high-multiplicity triggers are used to select the HM events. The first trigger uses the  VZERO-A and VZERO-C detectors~\cite{Abbas:2013taa} to estimate the multiplicity.  These are scintillator arrays located on both sides of the interaction point and covering a pseudorapidity range of $-3.7<\eta<-1.7$ and $2.8<\eta< 5.1$. They provide a signal whose amplitude is proportional to the number of charged particles. The second HM trigger is provided by the Silicon Pixel Detector (SPD) which comprises the two innermost layers of the Inner Tracking System (ITS)~\cite{Aamodt:2008zz}. The estimation of the multiplicity is based on the number of Fast-OR signals, which indicate the presence of at least one hit in the 
SPD. The total number of high-multiplicity triggered events which are  used in this analysis is $1.23 \times 10^{9}$.

The HNU events are selected using a dedicated trigger on nuclei provided by the TRD, which is described in detail in~\cite{ALICE:2017ymw}. The HNU trigger is a hardware-based single-track trigger, developed to enhance the sample of Z=2 nuclei (\he\ and $\alpha$) by selecting tracks with a high charge deposit in the TRD gas volume. The used HNU events were collected during the pp campaigns in 2017 and 2018, where the HNU trigger inspected $9.82\times10^{10}$ events, triggering $56.4\times10^6$ events. The inspected events satisfy the minimum-bias trigger condition, i.e., there is at least one hit in VZERO-A and VZERO-C. 
We refer to the event sample selected with the HNU trigger as the minimum-bias (MB) sample.

In both data samples, only events with a single reconstructed primary vertex are accepted, in order to reject collision pileup. The position of the primary vertex along the beam direction must be less than 10 cm from the nominal interaction point at the center of the experiment to ensure full geometrical acceptance in the ITS.

%%%%%%%%%%%%%%%%%%%%%%%%%% ANALYSIS TECHNIQUE %%%%%%%%%%%%%%%%%%%%%%%%%%%%%%%%%%%%%%%%%%%%%%%%%%%%%%%
%\section{Analysis Technique}
\subsection*{Hypertriton reconstruction}
For the identification of the $^{3}$He and $\pi$ candidates, information on the specific  energy loss (d$E$/d$x$)
 from  the ionization of the gas in the Time Projection Chamber (TPC) is used. 
The measured d$E$/d$x$ of a track is compared to the expected value from the Bethe-Bloch function, which describes the expected energy loss in the TPC as a function of the particle's velocity, for a given particle species and momentum. Tracks with a measured dE/dx within 3$\sigma$ of the expected Bethe-Bloch value are identified as that particle species. The resolution $\sigma$ of the specific energy loss measured in the TPC is about 6$\%$~\cite{Abelev:2014ffa}.

The hypertriton secondary vertex is identified using the V$^0$-finder algorithm~\cite{Abelev:2014ffa}, which reconstructs the decay of a neutral parent into two charged daughters (V$^0$ decay) during the data reconstruction stage. The algorithm uses the local properties of the helices describing the daughter trajectories, allowing it to account for the material budget in the reconstruction of the charged tracks.

The rapidity range of the  \hyp\ and \ahyp\ candidates is set to \mbox{$\mid y \mid < 0.8$} for the HM sample and \mbox{$\mid y \mid < 0.5$} for the MB sample. Additional topological and kinematic cuts, studied in Monte Carlo simulations, are applied to the data to reduce combinatorial background and improve signal extraction from the invariant mass distributions of \hyp\ and \ahyp\ candidates. The cuts are applied to the decay length ($ct$) , the distance-of-closest-approach  ($DCA$) between the $^{3}$He and $\pi$ daughter tracks, the $DCA$ between the \he\ track and the primary vertex, the transverse momentum of the daughters, and the cosine of the pointing angle (cos($\theta_{\mathrm{pointing}}$)), which is the angle between the reconstructed hypertriton momentum vector and the line connecting the primary and secondary vertices. Some of these selection criteria are strongly \pt-dependent, such as the trigger efficiency of the TRD nuclei trigger. For this reason, the optimization of the cuts is performed separately for the HM and MB samples.
 The selection criteria used for both data samples are listed in Table~\ref{cut_tab}.

\begin{table}[!htb]
\centering
\begin{tabular}{|c|c|c|}
\hline
\textbf{Selection criteria} & \textbf{High multiplicity sample} & \textbf{Minimum bias sample} \\
\hline
\hline
$\mid y \mid$ & $<$ 0.8 & $<$ 0.5 \\
\hline
$ct$ & $<$ 50 cm & $<$ 50 cm \\
\hline
cos($\theta_{\mathrm{pointing}}$) & $>$ 0.9995 & $>$ 0.998 \\
\hline
$DCA_{\mathrm{tracks}}$ & $<$ 1.5 cm  & $<$ 0.2 cm \\
\hline
$^3$He $DCA_{\mathrm{toPV}}$ & none & $<$ 2 cm \\
\hline
$\pi$ \pt\ & none & $>$ 0.1 GeV/\textit{c} \\
\hline
$^3$He \pt\ & $>$ 1.6 GeV/\textit{c}  & none \\
\hline
\end{tabular}
\caption{\hyp\ + \ahyp\  selection criteria}
\label{cut_tab}
\end{table}

\subsection*{Raw-yield extraction}
\label{raw_yield}

\begin{figure}[!htb]
\centering
\includegraphics[scale=0.38]{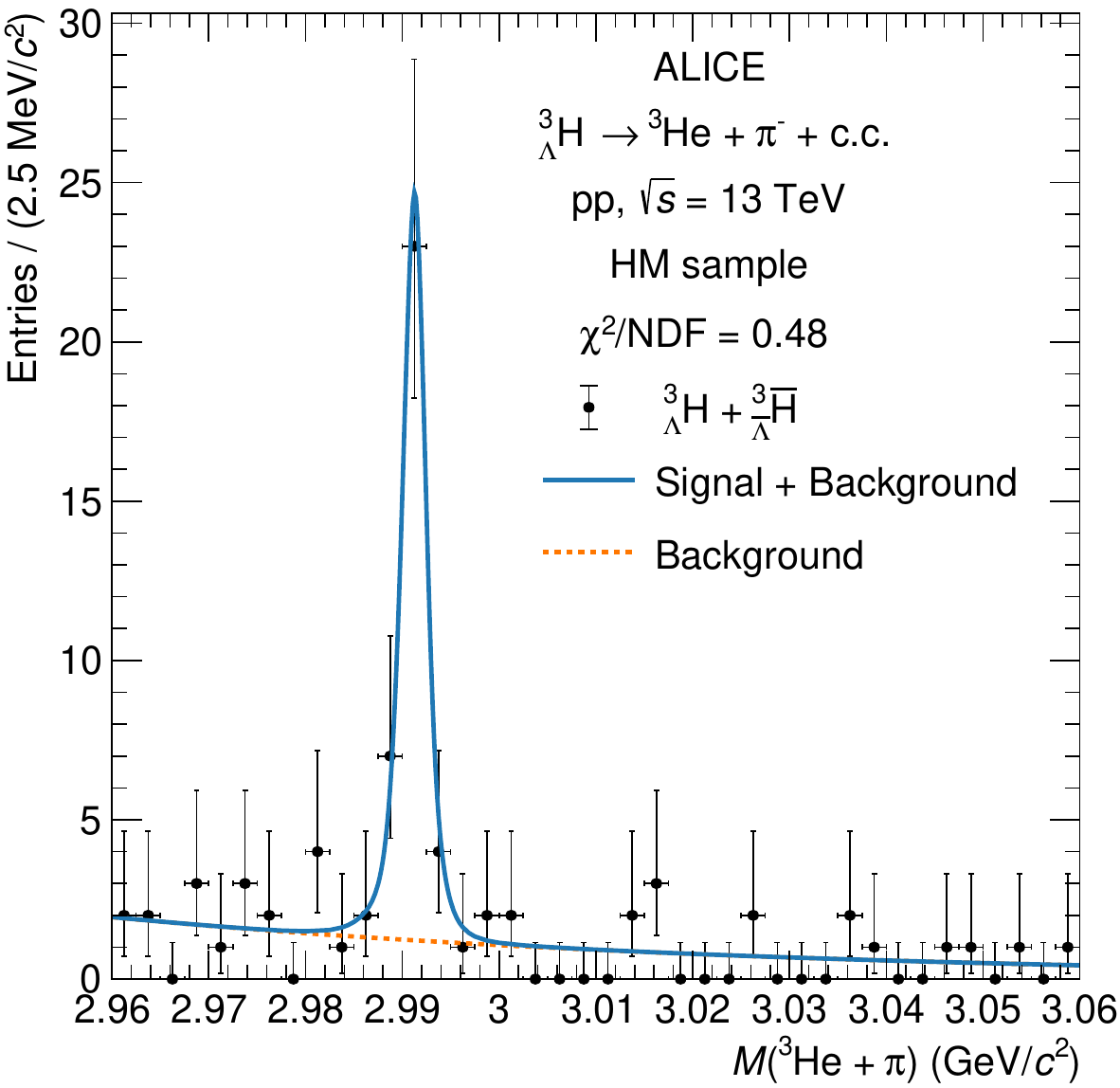}
\includegraphics[scale=0.38]{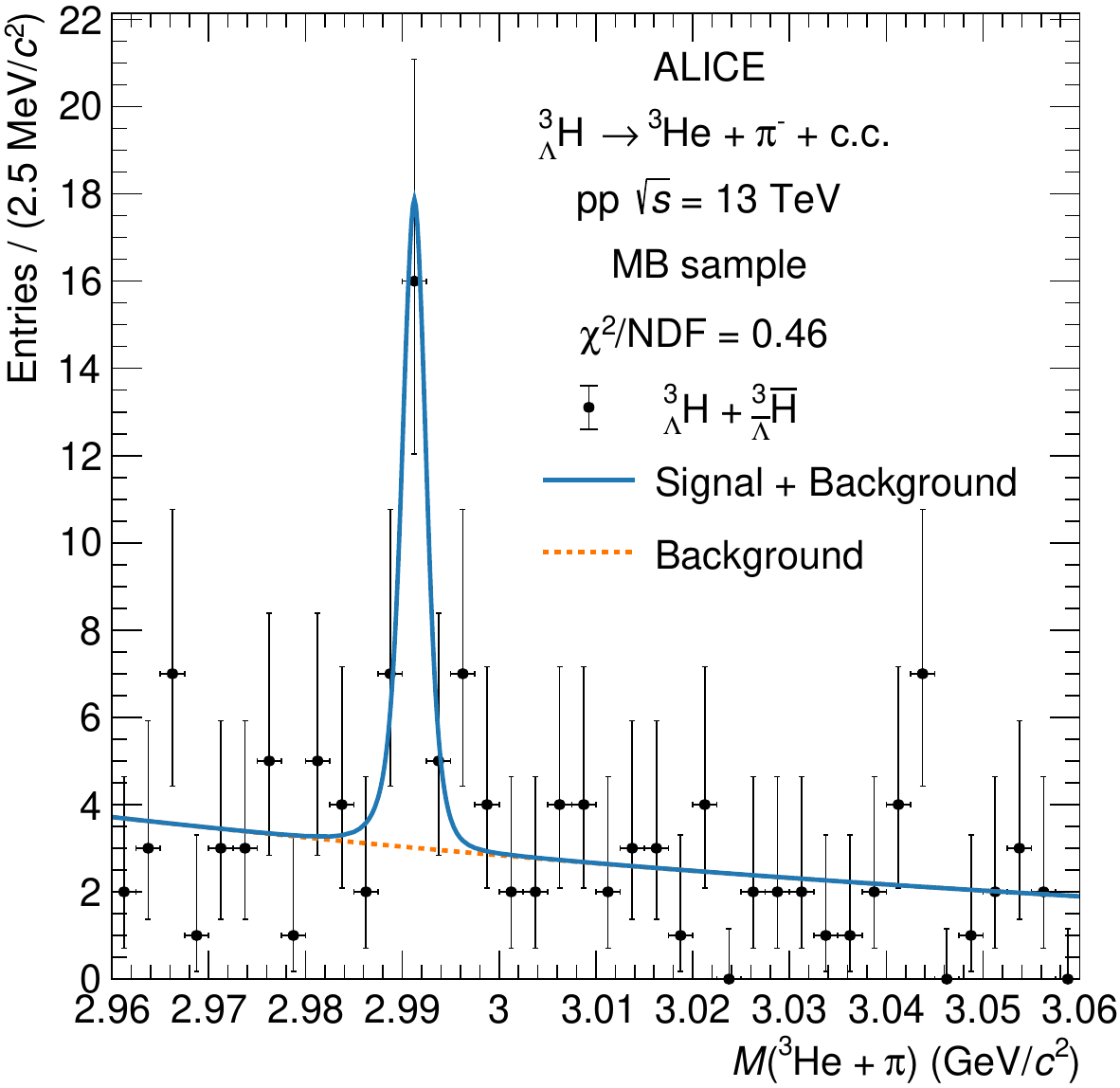}
\caption{ Invariant-mass spectra of (\hyp\ + \ahyp) integrated over \pt,  for the HM data sample (left) and the MB sample (right). The fitted functions used to extract the raw yields are shown as blue curves, while the background components are shown as orange dashed curves.}
\label{mass_inv_TRD}
\end{figure}

The raw yield is extracted by fitting the invariant-mass distribution ($\pi^-$+ $^3$He and  $\pi^+$+ $^3\overline{\rm He}$) with a likelihood fit. The fit function contains a signal and a background component. The background component is modeled by an exponential function. The signal component is taken from a template extrapolated from a Monte Carlo signal distribution and smoothed using a Kernel Density Estimation function (KDE). The fitted mass spectra are shown in Fig.~\ref{mass_inv_TRD} for both data samples, where no selection on the transverse momentum is applied. The high significance of the signal in the HM sample ($9.5\,\sigma$) allows a raw yield extraction in two separate \pt 
intervals ([1.6 -- 3.5] \GeVc and [3.5 -- 7] \GeVc), where the significance in each bin is at least 3. Below 1.6 \GeVc no candidates have been reconstructed. 

%%%%%%%%%%%%%%%%%%% Efficiency
\subsection*{Acceptance and efficiency corrections}
\label{eff_vs_ct_TRD}
The measured raw yields are modified by the finite kinematic acceptance of the ALICE detectors and the reconstruction efficiency. The acceptance is determined by the geometric coverage of the detectors, while the efficiency depends on detector conditions, the reconstruction algorithm, the selection criteria, and other factors that can affect track reconstruction. In addition, the selection efficiency of the HNU trigger is strongly \pt dependent which must be taken into account.

Corrections for the finite acceptance and efficiency are performed using Monte Carlo simulations that incorporate the detector geometry and the actual data-taking conditions. For the correction of the raw yields, the product \mbox{efficiency $\times$ acceptance} is evaluated as a function of \pt:

\begin{equation}
\textrm{acceptance} \times \textrm{efficiency} = \dfrac{N^{\rm rec}(\pt)}{N^{\rm gen}(\pt)},
\end{equation}
where $N^{\rm rec}$ is the number of reconstructed true \hyp\ (\ahyp)  that satisfy the criteria summarized in Table~\ref{cut_tab} and  $N^{\rm gen}$ is the number of generated \hyp\ (\ahyp) in the same rapidity range. 
Moreover, the same PID criteria for the daughter tracks as in data are required for the evaluation of $N^{\rm rec}$. For the HNU trigger, the reconstructed \hyp\ (\ahyp) must fulfill the HNU trigger conditions, while the generated ones must satisfy the MB trigger conditions. With this definition the resulting acceptance $\times$ efficiency includes also the HNU trigger efficiency. The different trigger efficiency for particles and antiparticles requires a separate correction. Thus the acceptance $\times$ efficiency is calculated separately for \hyp\ and \ahyp. 

%%%%%%%%%%%%%%%%%%%%%%%%%%%%%%%%  Integrated yield  %%%%%%%%%%%%%%%%%%%%%%%%%%%%%%%%%%%%%%%%%%%%%%%%%

\subsection*{Determination of the rapidity density $\mathrm{d}N/\mathrm{d}y$}

\subsubsection*{Minimum-bias sample}

For the HNU triggered data set, the significance is not sufficient to extract the \hyp\  and \ahyp\ signals in more than one \pt interval. The rapidity densities are computed as

\begin{equation}
\mathrm{d}N/\mathrm{d}y = \frac{1}{N_{\rm{ev}}\cdot {\mathrm{d}y}\cdot{\epsilon}\cdot{\rm B.R.}\cdot{f_{\rm abs}}}\cdot N_{\rm raw},
\label{corr_yield}
\end{equation}

where $N_{\rm ev}$ is the number of inspected events, the branching ratio (B.R.) is assumed to be 25\% and $f_{\rm abs}=0.97$ is the correction on the absorption. $\epsilon$ is the mean efficiency, i.e. the convolution of acceptance $\times$ efficiency 
weighted with the L\'evy-Tsallis \pt distribution obtained from the ALICE \he\ analysis~\cite{ALICE:2021mfm}. The yields are calculated separately for \hyp\ and \ahyp\ leading to an integrated yield of [2.4 $\pm$ 1.0 (stat.)]$\times10^{-8}$ for \hyp\ and [1.9 $\pm$ 0.7 (stat.)]$\times10^{-8}$ for \ahyp. The rapidity density is calculated as the average of the integrated \hyp\ and \ahyp\ yields:
$${\mathrm{d}N}/{\mathrm{d}y}^{\mathrm {MB}} = \left[2.1 \pm 0.6\, \mathrm{(stat.)} \pm 0.4\, \mathrm{(syst.)} \right] \times 10^{-8}.$$ 

\subsubsection*{High-multiplicity sample}
\begin{figure}[!htb]
\centering
\includegraphics[scale=0.5]{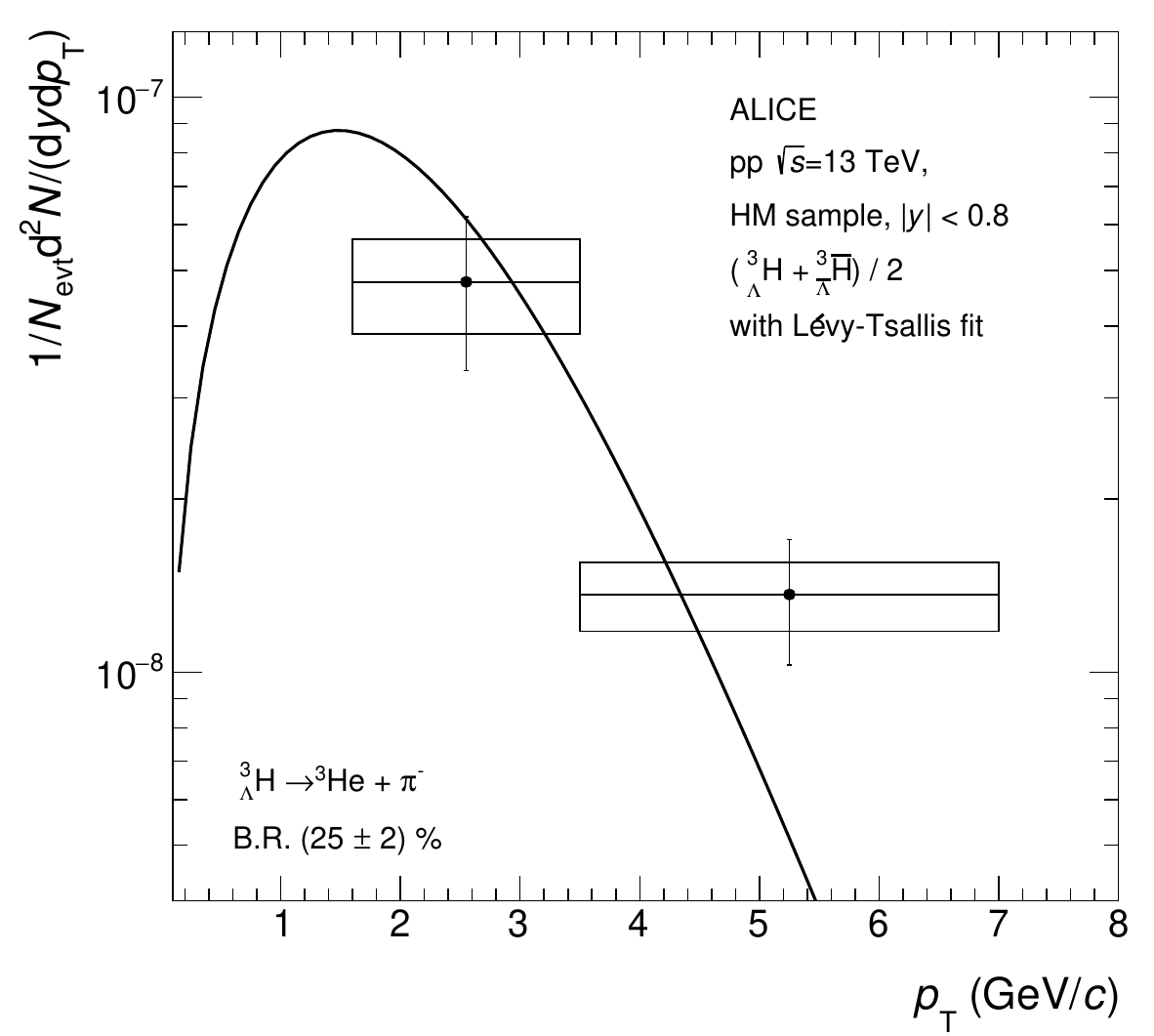}
\caption{Corrected \pt\ spectra of (\hyp\ + \ahyp)/2 in the HM data sample, fitted with a L\'evy-Tsallis function.}
\label{corrected_spectrum_HM}
\end{figure}

The fully corrected \pt spectrum of (\hyp$+$\ahyp)$/2$ is shown in Fig.~\ref{corrected_spectrum_HM}, where the yield in each \pt bin is calculated using Eq.~(\ref{corr_yield}). To obtain 
the rapidity density $\mathrm{d}N/\mathrm{d}y$, the spectrum is fitted with a L\'evy-Tsallis function. The 
shape of the fit function is constrained by the parameters of a fit of the  L\'evy-Tsallis function to the  \he\ spectrum from the same data set~\cite{ALICE:2021mfm}, with the normalization treated as a free parameter. The resulting \hyp\ rapidity density and its statistical and systematic uncertainties are:

$$\mathrm{d}N/\mathrm{d}y^{\mathrm {HM}} = \left( 2.4 \pm 0.5\, \mathrm{stat.} \pm 0.3\, \mathrm{syst.} \right) \times 10^{-7}.$$

The determination of systematic uncertainties is described in the next section.

%%%%%%%%%% SYSTEMATICS
\subsection*{Systematic uncertainties}
Systematic uncertainties on the corrected yields arise from various sources. 
A detailed study was performed where the different contributions to the systematic
uncertainty were determined by a variation of the applied selection criteria, assumptions, 
and methods:
\begin{itemize}

\item {\em Particle identification \& Tracking}: The number of standard deviations for the particle identification (PID) and the track selection criteria on the daughter tracks are varied.

\item {\em Topological selection criteria}: The topological and kinematic constraints to select the (anti)hypertriton candidates are varied.

\item {\em Absorption}: An absorption correction of 3\% is applied with a systematic uncertainty of 4\% as reported in~\cite{ALargeIonColliderExperiment:2021puh}.

\item {\em Signal extraction}: Different fit functions are used to fit the invariant-mass distributions. For the signal component a double-sided Crystal Ball (Gaussian with tails on both sides) fit is used instead of the KDE and for the background, polynomials of zero, first and second order are used instead of the exponential function.

\item {\em Branching ratio}: The branching ratio of the \hyp\, (\ahyp) decay into $^{3}$He and $\pi^{-}$  is assumed to be ($25\pm2$)\,\% as reported in~\cite{ALargeIonColliderExperiment:2021puh}, which corresponds to a systematic uncertainty of 9\% for the yields.
The uncertainty is estimated as the difference between the theoretical value~\cite{Kamada:1997rv}, which is also used as the central value, and the measurement by the STAR Collabboration~\cite{STAR:2017gxa}.

\item {\em Extrapolation to $\pt=0$}: The assumption of the shape of the \pt spectrum gives rise to a systematic uncertainty which is estimated by using different functional forms. To this end, Boltzmann, $m_T$- and $p_{\rm{T}}$-exponential, Boltzmann-Gibbs Blast-Wave and Fermi-Dirac functions are used instead of the default L\'evy-Tsallis function.

\item {\em Material budget}: The uncertainty of the material budget in the Monte Carlo simulations is 4.5\%~\cite{Abelev:2014ffa} which results in an uncertainty of 2\% in the yields~\cite{ALICE:2021mfm}.

\end{itemize}
In each category where variations are performed, the difference between the uppermost and the lowermost yield is divided by $\sqrt{12}$ to obtain the uncertainty. The resulting uncertainties for all categories are listed in Tab.~\ref{syst_tab} The total uncertainty is the quadratic sum of the individual contributions. 

\begin{table}[!htb]
\centering
\begin{tabular}{|c|c|c|}
\hline
Category & Syst. uncertainty on $\mathrm{d}N/\mathrm{d}y^{\mathrm {MB}}$ & Syst. uncertainty on $\mathrm{d}N/\mathrm{d}y^{\mathrm {HM}}$ \\
\hline
\hline
PID \& Tracking & 9\,\% &  6\,\% \\
\hline
Topological cuts & 11\,\% &  4\,\% \\
\hline
Absorption & 4\,\% &  4\,\% \\
\hline
Signal extraction & 1\,\% &  4\,\% \\
\hline
Branching ratio & 9\,\% & 9\,\% \\
\hline
Extrapolation to \pt = 0& 7\,\% &  8\,\% \\
\hline
Material budget & 2\,\% &  2\,\% \\
\hline
Total & 19\,\% &  15\,\% \\
\hline
\end{tabular}
\caption{Systematic uncertainties on the yield for each category and both data samples. Furthermore, the total uncertainty is shown, calculated as the quadratic sum of the individual contributions. }
\label{syst_tab}
\end{table}

%%%%%%%%%%%%%%%%%%%%%% hypertriton radius %%%%%%%%%%%%%%%%%%%%%%%%%%%%%%%%%%%%%%%%%%%%%%%%%%%%%%%%%%%%%%
\subsection*{Determination of the hypertriton radius}

In the analytical coalescence approach~\cite{Sun:2018mqq}, the suppression of nuclear cluster production in small collision systems is related to the ratio of the cluster size to the size of the nucleon-emitting source. This provides a new tool for investigating the nuclear wave function by measuring the nuclear production yields, provided the source size is known (which will be discussed below).

In the following, we briefly outline the arguments presented in~\cite{Sun:2018mqq}. Before applying this technique to determine the radius of the hypertriton, we use it to extract the radius of objects of known size, i.e. the deuteron and the \he\ nucleus, and compare the results with the literature values.

\subsubsection*{Fit of the deuteron radius}
In the analytical coalescence model~\cite{Sun:2018mqq}, the d/p ratio in a collision system with the size $R$ of the nucleon-emitting source is given by:

\begin{equation}
\label{coal_Deuteron}
\mathrm{d/p}= \frac{C_1}{\left[ 1 + \left( \frac{\sigma}{2R} \right)^2 \right]^{3/2}},
\end{equation}
where $\sigma$ is the Gaussian size parameter of the deuteron's Wigner function, assuming a harmonic-oscillator wave function.
Its is related to the deuteron root-mean-square matter radius \rd by 
\begin{equation}
\label{sigma_rd}
\sigma=\sqrt{\frac{8}{3}}\cdot \rd. 
\end{equation}
The parameter  $C_1$ represents the limit  of d/p for $R\gg\sigma$ and is 
experimentally determined by the d/p value measured in Pb--Pb collisions, leading to $C_1= [4.0 \pm 0.2]\times 10^{-3}$~\cite{ALICE:2022veq}. The charged-particle dependence of the nucleon source size $R$ in~\cite{Sun:2018mqq} follows a theoretical approach where it is connected to the production rate of protons that is parameterized as a function of $\langle \dndeta \rangle$ based on measurements from ALICE~\cite{ALICE:2015tra,Abelev:2013vea,Adam:2015qaa}.

A data-driven approach is used to determine object sizes. The aim is to determine the source size evolution as a function of multiplicity exploiting measurements in pp and Pb--Pb collisions.
The size of the particle-emitting source can be determined by analyzing the momentum correlations of particle pairs. In a study by ALICE~\cite{ALICE:2023sjd} focusing on high-multiplicity \pp\ collisions, the primordial source size was evaluated as a function of the transverse mass ($m_{\rm T}$) of the pairs. The results indicate a common emission source for all hadrons in small collision systems at the LHC. This observation enables the use of source size measurements derived from $\pi\pi$ correlation studies in minimum bias \pp\ collisions~\cite{ALICE:2020ibs,ALICE:2023zbh,ALICE:2025aur}, which were performed in three different charged-particle multiplicity classes. In detail, the model needs the common source of the nucleons, i.e. protons in our case, as input and pions are only an approximation. This is connected to different resonances feeding down into the pions and the protons. (It was tested in our case to be less problematic, since other factors give larger contribution.)
In addition, we include one measurement from proton–proton correlation data in high-multiplicity \pp\ collisions~\cite{ALICE:2023sjd}, along with three measurements from proton–proton correlations in central, semi-central, and peripheral \PbPb\ collisions~\cite{ALICE:2015hvw,ALICE:2025wuy}. In all cases, a transverse mass of \mt = 1.4 GeV$/{c}^{2}$ is used. If necessary, measurements are interpolated to this value. The choice of 
\mt = 1.4 GeV$/{c}^{2}$ corresponds to the average coalescence momentum 
relevant for our analysis of about \pt = 1 GeV$/c$ with a nucleon mass of about 1 GeV$/{c}^{2}$.

All correlation-based source sizes are extracted assuming a one-dimensional Gaussian distribution. However, the analytical coalescence model adopted in this study assumes a three-dimensional Gaussian source. Therefore, the measured one-dimensional source radii are corrected by a factor of 2.25 to account for this dimensional difference.

The dependence of the source size $R$ on the average charged-particle density \avdndeta is parametrized by fitting the following functional form to the extracted source sizes:

\begin{equation}
\label{sourcePara}
R(\avdndeta)=\alpha + \beta \times \avdndeta^{\gamma}.
\end{equation}

From the fit we obtain $\alpha = 0.40 \pm 0.17$, $\beta = 0.62 \pm 0.27$, and $\gamma = 0.38 \pm 0.05$. The statistical uncertainty of $R(\avdndeta)$ is derived from the fit. To estimate the systematic uncertainty, all source size values are uniformly shifted up and down by one standard deviation of their respective systematic uncertainties, and the fit is repeated. The envelope of the resulting fit variations defines the systematic uncertainty on the parametrization.

With the source size determined, the radius of the coalescing object can be calculated using the analytical coalescence model. Equation~\ref{coal_Deuteron}  is fitted to the measured d/p ratio as a function of \avdndeta with $\sigma$ as a free parameter. The deuteron radius \rd and the corresponding statistical uncertainty are derived from the fit, where relation~\ref{sigma_rd} is used. This is done individually for each measured d/p ratio. To obtain the combined radius, the fit is applied simultaneously to all d/p ratios. 

Furthermore, we examined the systematic uncertainties. The values of $C_1$ and $R$ are shifted up and down by one standard deviation and the fit is performed again. The systematic uncertainty is calculated as the difference between the minimum and maximum radius. This procedure is done separately for  $C_1$ and $R$. The resulting systematic uncertainty is the squared sum of both contributions.

\subsubsection*{Fit of the \he\  and \hyp\  radii  }
The \he\ radius is determined using the three-body coalescence model for the \he/p ratio:

\begin{equation}
\label{coal3_He}
\mathrm{^3He/p} = \frac{C_2}{\left[ 1 + \biggl(\frac{\rhe}{\sqrt{2}R} \biggr)^2 \right]^{3}}
\end{equation}
where $C_2 = 4C_1^2 / 9 = [7.1\pm0.7]\times 10^{-6}$. The \he\ matter radius \rhe and its uncertainties are determined following the same procedure as for the deuteron.

The \hyp/$\Lambda$ coalescence formula (Eq.~\ref{coal2_Hyp}) also contains $C_2$. However, there is an additional correction factor (0.85) that takes into account the mass difference between nucleons and the $\Lambda$ hyperon. The factor $\sqrt{2/9}$ in~\ref{coal2_Hyp} originates from the coordinate transformation into the d--$\Lambda$ system, which is required for two-body coalescence. The central values of the object size and its uncertainties are determined in the same way as described for the deuteron radius. 

Further studies have been performed to address uncertainties related to the coalescence model and the Gaussian wave function used in this analysis. A non-analytical coalescence model, described in~\cite{Mahlein:2025bla}, is applied for this purpose. This model allows the usage of various wave functions, but is limited to pp collisions. This model is used to determine the radii in a similar way, but only the two measured \hyp/$\Lambda$ ratios from pp collisions are included. This results in an object size of $11.03^{+0.56}_{-0.56}$ fm with a Gaussian wave function and $13.03^{+2.41}_{-1.53}$ fm with a Congleton wave function~\cite{Congleton:1992kk}. To obtain comparable values, the object size is determined using the analytical coalescence model with the same \hyp/$\Lambda$ ratios, resulting in 9.83 fm. The difference between the object size determined using the analytical and non-analytical models when using the same wave function (Gaussian) results in a model uncertainty of ${}^{+1.20}_{-0.00}$ fm. The difference between the results of the non-analytical model with different wave functions (Gaussian/Congleton) yields an uncertainty with respect to the wave function of ${}^{+2.00}_{-0.00}$ fm. Both uncertainties are added quadratically to the previously determined uncertainties.

In addition, we determined the size of the hypertriton using three-body coalescence, following the same fitting procedure as described above. We obtain a matter radius of  \rhyp = $4.26^{+0.38}_{-0.33}$\, fm, where the uncertainties take into account all statistical and systematic uncertainties related to the measured yield ratios and the source size. Using the coordinate transformation described above, we obtain the corresponding d-$\Lambda$ distance of $\rrhyp = 9.03^{+0.81}_{-0.70}$ fm.

We also checked the influence on the object if the source size is determined from \avdndeta using the theoretical approach as described in~\cite{Sun:2018mqq}. We use the same methods to determine the object sizes and their uncertainties as outlined above and we obtain \rd = $1.87^{+0.18}_{-.018}$\,fm, \rhe = $2.20^{+0.28}_{-0.29}$\,fm, and \rrhyp = $9.81^{+2.72}_{-1.04}$\,fm, which are in good agreement with the results obtained with the data-driven approach.

\subsection*{Determination of $B_{\Lambda}$}
The distance between deuteron and $\Lambda$ in the hypertriton can be calculated from the $\Lambda$ separation energy using pionless effective field theory~\cite{Hildenbrand:2019sgp}, shown as the blue band in Fig.~\ref{fig:B_Lambda}. In addition, the prediction of a simple quantum mechanical model~\cite{Braun-Munzinger:2018hat} is shown in Fig.~\ref{fig:B_Lambda} as a dashed black curve, which is in agreement with the EFT model. We use the measured hypertriton size, represented by the red band in Fig.~\ref{fig:B_Lambda}, to determine $B_\Lambda$ from the intersection of the central values of the measured radius and the EFT predictions. The uncertainty of the EFT prediction is a naive estimate of the uncertainty due to the two-pion exchange in the $\Lambda$--p interaction~\cite{Hildenbrand:2019sgp}. The uncertainty of the hypertriton size is the statistical and systematic uncertainty added in quadrature. In both cases, a Gaussian distribution is assumed. To determine the uncertainty of $B_\Lambda$, both bands are considered as probability density functions and multiplied. The resulting model is then fitted to a toy data point set at the intersection of the central values. The $1\sigma$ region is then calculated from the two-dimensional fit and shown as a red contour in Fig.~\ref{fig:B_Lambda}. The maximum extent of the contour in the $B_\Lambda$ direction is used as total uncertainty.

\clearpage

%%%%% Authorlist - please do not touch: handled by EB chairs 
\section{The ALICE Collaboration}
\label{app:collab}
% ALICE Collaboration author list for 2026-02-13
\begin{flushleft} 
\small

D.A.H.~Abdallah\,\orcidlink{0000-0003-4768-2718}\,$^{\rm 134}$, 
I.J.~Abualrob\,\orcidlink{0009-0005-3519-5631}\,$^{\rm 112}$, 
S.~Acharya\,\orcidlink{0000-0002-9213-5329}\,$^{\rm 49}$, 
K.~Agarwal\,\orcidlink{0000-0001-5781-3393}\,$^{\rm II,}$$^{\rm 23}$, 
G.~Aglieri Rinella\,\orcidlink{0000-0002-9611-3696}\,$^{\rm 32}$, 
L.~Aglietta\,\orcidlink{0009-0003-0763-6802}\,$^{\rm 24}$, 
N.~Agrawal\,\orcidlink{0000-0003-0348-9836}\,$^{\rm 25}$, 
Z.~Ahammed\,\orcidlink{0000-0001-5241-7412}\,$^{\rm 132}$, 
S.~Ahmad\,\orcidlink{0000-0003-0497-5705}\,$^{\rm 15}$, 
I.~Ahuja\,\orcidlink{0000-0002-4417-1392}\,$^{\rm 36}$
% ...

\section*{Affiliation Notes}

$^{\rm I}$ Deceased\\
$^{\rm II}$ Also at: INFN Trieste\\
$^{\rm III}$ Also at: Fondazione Bruno Kessler (FBK), Trento, Italy\\
$^{\rm IV}$ Also at: Max-Planck-Institut fur Physik, Munich, Germany\\
$^{\rm V}$ Also at: Czech Technical University in Prague (CZ)\\
$^{\rm VI}$ Also at: Instituto de Fisica da Universidade de Sao Paulo\\
$^{\rm VII}$ Also at: Dipartimento DET del Politecnico di Torino, Turin, Italy\\
$^{\rm VIII}$ Also at: Department of Applied Physics, Aligarh Muslim University, Aligarh, India\\
$^{\rm IX}$ Also at: Institute of Theoretical Physics, University of Wroclaw, Poland\\
$^{\rm X}$ Also at: Facultad de Ciencias, Universidad Nacional Aut\'{o}noma de M\'{e}xico, Mexico City, Mexico\\

\section*{Collaboration Institutes}

$^{1}$ A.I. Alikhanyan National Science Laboratory (Yerevan Physics Institute) Foundation, Yerevan, Armenia\\
$^{2}$ AGH University of Krakow, Cracow, Poland\\
$^{3}$ Bogolyubov Institute for Theoretical Physics, National Academy of Sciences of Ukraine, Kyiv, Ukraine\\
$^{4}$ Bose Institute, Department of Physics  and Centre for Astroparticle Physics and Space Science (CAPSS), Kolkata, India\\
$^{5}$ California Polytechnic State University, San Luis Obispo, California, United States\\
$^{6}$ Central China Normal University, Wuhan, China\\
$^{7}$ Centro de Aplicaciones Tecnol\'{o}gicas y Desarrollo Nuclear (CEADEN), Havana, Cuba\\
$^{8}$ Centro de Investigaci\'{o}n y de Estudios Avanzados (CINVESTAV), Mexico City and M\'{e}rida, Mexico\\
$^{9}$ Chicago State University, Chicago, Illinois, United States\\
$^{10}$ China Nuclear Data Center, China Institute of Atomic Energy, Beijing, China\\
$^{11}$ China University of Geosciences, Wuhan, China\\
$^{12}$ Chungbuk National University, Cheongju, Republic of Korea\\
$^{13}$ Comenius University Bratislava, Faculty of Mathematics, Physics and Informatics, Bratislava, Slovak Republic\\
$^{14}$ Creighton University, Omaha, Nebraska, United States\\
$^{15}$ Department of Physics, Aligarh Muslim University, Aligarh, India\\
$^{16}$ Department of Physics, Pusan National University, Pusan, Republic of Korea\\
$^{17}$ Department of Physics, Sejong University, Seoul, Republic of Korea\\
$^{18}$ Department of Physics, University of California, Berkeley, California, United States\\
$^{19}$ Department of Physics, University of Oslo, Oslo, Norway\\
$^{20}$ Department of Physics and Technology, University of Bergen, Bergen, Norway\\
$^{21}$ Dipartimento di Fisica, Universit\`{a} di Pavia, Pavia, Italy\\
$^{22}$ Dipartimento di Fisica dell'Universit\`{a} and Sezione INFN, Cagliari, Italy\\
$^{23}$ Dipartimento di Fisica dell'Universit\`{a} and Sezione INFN, Trieste, Italy\\
$^{24}$ Dipartimento di Fisica dell'Universit\`{a} and Sezione INFN, Turin, Italy\\
$^{25}$ Dipartimento di Fisica e Astronomia dell'Universit\`{a} and Sezione INFN, Bologna, Italy\\
$^{26}$ Dipartimento di Fisica e Astronomia dell'Universit\`{a} and Sezione INFN, Catania, Italy\\
$^{27}$ Dipartimento di Fisica e Astronomia dell'Universit\`{a} and Sezione INFN, Padova, Italy\\
$^{28}$ Dipartimento di Fisica `E.R.~Caianiello' dell'Universit\`{a} and Gruppo Collegato INFN, Salerno, Italy\\
$^{29}$ Dipartimento DISAT del Politecnico and Sezione INFN, Turin, Italy\\
$^{30}$ Dipartimento di Scienze MIFT, Universit\`{a} di Messina, Messina, Italy\\
$^{31}$ Dipartimento Interateneo di Fisica `M.~Merlin' and Sezione INFN, Bari, Italy\\
$^{32}$ European Organization for Nuclear Research (CERN), Geneva, Switzerland\\
$^{33}$ Faculty of Electrical Engineering, Mechanical Engineering and Naval Architecture, University of Split, Split, Croatia\\
$^{34}$ Faculty of Nuclear Sciences and Physical Engineering, Czech Technical University in Prague, Prague, Czech Republic\\
$^{35}$ Faculty of Physics, Sofia University, Sofia, Bulgaria\\
$^{36}$ Faculty of Science, P.J.~\v{S}af\'{a}rik University, Ko\v{s}ice, Slovak Republic\\
$^{37}$ Faculty of Technology, Environmental and Social Sciences, Bergen, Norway\\
$^{38}$ Frankfurt Institute for Advanced Studies, Johann Wolfgang Goethe-Universit\"{a}t Frankfurt, Frankfurt, Germany\\
$^{39}$ Fudan University, Shanghai, China\\
$^{40}$ Gauhati University, Department of Physics, Guwahati, India\\
$^{41}$ Helmholtz-Institut f\"{u}r Strahlen- und Kernphysik, Rheinische Friedrich-Wilhelms-Universit\"{a}t Bonn, Bonn, Germany\\
$^{42}$ Helsinki Institute of Physics (HIP), Helsinki, Finland\\
$^{43}$ High Energy Physics Group,  Universidad Aut\'{o}noma de Puebla, Puebla, Mexico\\
$^{44}$ Horia Hulubei National Institute of Physics and Nuclear Engineering, Bucharest, Romania\\
$^{45}$ HUN-REN Wigner Research Centre for Physics, Budapest, Hungary\\
$^{46}$ Indian Institute of Technology Bombay (IIT), Mumbai, India\\
$^{47}$ Indian Institute of Technology Indore, Indore, India\\
$^{48}$ INFN, Laboratori Nazionali di Frascati, Frascati, Italy\\
$^{49}$ INFN, Sezione di Bari, Bari, Italy\\
$^{50}$ INFN, Sezione di Bologna, Bologna, Italy\\
$^{51}$ INFN, Sezione di Cagliari, Cagliari, Italy\\
$^{52}$ INFN, Sezione di Catania, Catania, Italy\\
$^{53}$ INFN, Sezione di Padova, Padova, Italy\\
$^{54}$ INFN, Sezione di Pavia, Pavia, Italy\\
$^{55}$ INFN, Sezione di Torino, Turin, Italy\\
$^{56}$ INFN, Sezione di Trieste, Trieste, Italy\\
$^{57}$ Inha University, Incheon, Republic of Korea\\
$^{58}$ Institute for Gravitational and Subatomic Physics (GRASP), Utrecht University/Nikhef, Utrecht, Netherlands\\
$^{59}$ Institute of Experimental Physics, Slovak Academy of Sciences, Ko\v{s}ice, Slovak Republic\\
$^{60}$ Institute of Physics, Homi Bhabha National Institute, Bhubaneswar, India\\
$^{61}$ Institute of Physics of the Czech Academy of Sciences, Prague, Czech Republic\\
$^{62}$ Institute of Space Science (ISS), Bucharest, Romania\\
$^{63}$ Institut f\"{u}r Kernphysik, Johann Wolfgang Goethe-Universit\"{a}t Frankfurt, Frankfurt, Germany\\
$^{64}$ Instituto de Ciencias Nucleares, Universidad Nacional Aut\'{o}noma de M\'{e}xico, Mexico City, Mexico\\
$^{65}$ Instituto de F\'{i}sica, Universidade Federal do Rio Grande do Sul (UFRGS), Porto Alegre, Brazil\\
$^{66}$ Instituto de F\'{\i}sica, Universidad Nacional Aut\'{o}noma de M\'{e}xico, Mexico City, Mexico\\
$^{67}$ iThemba LABS, National Research Foundation, Somerset West, South Africa\\
$^{68}$ Jeonbuk National University, Jeonju, Republic of Korea\\
$^{69}$ Korea Institute of Science and Technology Information, Daejeon, Republic of Korea\\
$^{70}$ Laboratoire de Physique Subatomique et de Cosmologie, Universit\'{e} Grenoble-Alpes, CNRS-IN2P3, Grenoble, France\\
$^{71}$ Lawrence Berkeley National Laboratory, Berkeley, California, United States\\
$^{72}$ Lund University Department of Physics, Division of Particle Physics, Lund, Sweden\\
$^{73}$ Marietta Blau Institute, Vienna, Austria\\
$^{74}$ Nagasaki Institute of Applied Science, Nagasaki, Japan\\
$^{75}$ Nara Women{'}s University (NWU), Nara, Japan\\
$^{76}$ National Centre for Nuclear Research, Warsaw, Poland\\
$^{77}$ National Institute of Science Education and Research, Homi Bhabha National Institute, Jatni, India\\
$^{78}$ National Nuclear Research Center, Baku, Azerbaijan\\
$^{79}$ National Research and Innovation Agency - BRIN, Jakarta, Indonesia\\
$^{80}$ Niels Bohr Institute, University of Copenhagen, Copenhagen, Denmark\\
$^{81}$ Nikhef, National institute for subatomic physics, Amsterdam, Netherlands\\
$^{82}$ Nuclear Physics Group, STFC Daresbury Laboratory, Daresbury, United Kingdom\\
$^{83}$ Nuclear Physics Institute of the Czech Academy of Sciences, Husinec-\v{R}e\v{z}, Czech Republic\\
$^{84}$ Oak Ridge National Laboratory, Oak Ridge, Tennessee, United States\\
$^{85}$ Ohio State University, Columbus, Ohio, United States\\
$^{86}$ Physics department, Faculty of science, University of Zagreb, Zagreb, Croatia\\
$^{87}$ Physics Department, Panjab University, Chandigarh, India\\
$^{88}$ Physics Department, University of Jammu, Jammu, India\\
$^{89}$ Physics Program and International Institute for Sustainability with Knotted Chiral Meta Matter (WPI-SKCM$^{2}$), Hiroshima University, Hiroshima, Japan\\
$^{90}$ Physikalisches Institut, Eberhard-Karls-Universit\"{a}t T\"{u}bingen, T\"{u}bingen, Germany\\
$^{91}$ Physikalisches Institut, Ruprecht-Karls-Universit\"{a}t Heidelberg, Heidelberg, Germany\\
$^{92}$ Physik Department, Technische Universit\"{a}t M\"{u}nchen, Munich, Germany\\
$^{93}$ Politecnico di Bari and Sezione INFN, Bari, Italy\\
$^{94}$ Research Division and ExtreMe Matter Institute EMMI, GSI Helmholtzzentrum f\"ur Schwerionenforschung GmbH, Darmstadt, Germany\\
$^{95}$ Saga University, Saga, Japan\\
$^{96}$ Saha Institute of Nuclear Physics, Homi Bhabha National Institute, Kolkata, India\\
$^{97}$ School of Physics and Astronomy, University of Birmingham, Birmingham, United Kingdom\\
$^{98}$ Secci\'{o}n F\'{\i}sica, Departamento de Ciencias, Pontificia Universidad Cat\'{o}lica del Per\'{u}, Lima, Peru\\
$^{99}$ SUBATECH, IMT Atlantique, Nantes Universit\'{e}, CNRS-IN2P3, Nantes, France\\
$^{100}$ Sungkyunkwan University, Suwon City, Republic of Korea\\
$^{101}$ Suranaree University of Technology, Nakhon Ratchasima, Thailand\\
$^{102}$ Technical University of Ko\v{s}ice, Ko\v{s}ice, Slovak Republic\\
$^{103}$ The Henryk Niewodniczanski Institute of Nuclear Physics, Polish Academy of Sciences, Cracow, Poland\\
$^{104}$ The University of Texas at Austin, Austin, Texas, United States\\
$^{105}$ Universidad Aut\'{o}noma de Sinaloa, Culiac\'{a}n, Mexico\\
$^{106}$ Universidade de S\~{a}o Paulo (USP), S\~{a}o Paulo, Brazil\\
$^{107}$ Universidade Estadual de Campinas (UNICAMP), Campinas, Brazil\\
$^{108}$ Universidade Federal do ABC, Santo Andre, Brazil\\
$^{109}$ Universitatea Nationala de Stiinta si Tehnologie Politehnica Bucuresti, Bucharest, Romania\\
$^{110}$ University of Cape Town, Cape Town, South Africa\\
$^{111}$ University of Derby, Derby, United Kingdom\\
$^{112}$ University of Houston, Houston, Texas, United States\\
$^{113}$ University of Jyv\"{a}skyl\"{a}, Jyv\"{a}skyl\"{a}, Finland\\
$^{114}$ University of Kansas, Lawrence, Kansas, United States\\
$^{115}$ University of Liverpool, Liverpool, United Kingdom\\
$^{116}$ University of Science and Technology of China, Hefei, China\\
$^{117}$ University of Silesia in Katowice, Katowice, Poland\\
$^{118}$ University of South-Eastern Norway, Kongsberg, Norway\\
$^{119}$ University of Tennessee, Knoxville, Tennessee, United States\\
$^{120}$ University of the Witwatersrand, Johannesburg, South Africa\\
$^{121}$ University of Tokyo, Tokyo, Japan\\
$^{122}$ University of Tsukuba, Tsukuba, Japan\\
$^{123}$ Universit\"{a}t M\"{u}nster, Institut f\"{u}r Kernphysik, M\"{u}nster, Germany\\
$^{124}$ Universit\'{e} Clermont Auvergne, CNRS/IN2P3, LPC, Clermont-Ferrand, France\\
$^{125}$ Universit\'{e} de Lyon, CNRS/IN2P3, Institut de Physique des 2 Infinis de Lyon, Lyon, France\\
$^{126}$ Universit\'{e} de Strasbourg, CNRS, IPHC UMR 7178, F-67000 Strasbourg, France, Strasbourg, France\\
$^{127}$ Universit\'{e} Paris-Saclay, Centre d'Etudes de Saclay (CEA), IRFU, D\'{e}partment de Physique Nucl\'{e}aire (DPhN), Saclay, France\\
$^{128}$ Universit\'{e}  Paris-Saclay, CNRS/IN2P3, IJCLab, Orsay, France\\
$^{129}$ Universit\`{a} degli Studi di Foggia, Foggia, Italy\\
$^{130}$ Universit\`{a} del Piemonte Orientale, Vercelli, Italy\\
$^{131}$ Universit\`{a} di Brescia, Brescia, Italy\\
$^{132}$ Variable Energy Cyclotron Centre, Homi Bhabha National Institute, Kolkata, India\\
$^{133}$ Warsaw University of Technology, Warsaw, Poland\\
$^{134}$ Wayne State University, Detroit, Michigan, United States\\
$^{135}$ Yale University, New Haven, Connecticut, United States\\
$^{136}$ Yildiz Technical University, Istanbul, Turkey\\
$^{137}$ Yonsei University, Seoul, Republic of Korea\\
$^{138}$ Affiliated with an institute formerly covered by a cooperation agreement with CERN\\
$^{139}$ Affiliated with an international laboratory covered by a cooperation agreement with CERN.\\

\end{flushleft} 

\end{document}